\documentclass{pasa}%

\usepackage{graphicx}
\usepackage{multicol}
\usepackage[titletoc,title]{appendix}
\usepackage{subcaption}
\usepackage[usenames,dvipsnames,svgnames,hyperref]{xcolor}
\usepackage{amsfonts}
\usepackage[
    pdftex,
    pdfpagelabels,
    breaklinks=true,
    bookmarks=true,
    bookmarksopen=false,
    bookmarksopenlevel=2
    bookmarksnumbered=true,
    bookmarkstype=toc,
    colorlinks=true,
    citecolor=RoyalBlue,
    linkcolor=ForestGreen,
    menucolor=Teal,
    urlcolor=DarkOrange,
]{hyperref}

\title[Merger Tree Dendograms]{Observing Merger Trees in a New Light}

\author[Poulton et al.]{Rhys J. J. Poulton$^{1,2,3}$, Aaron S.G. Robotham$^{1,2}$, Chris Power$^{1,2}$ and Pascal J. Elahi$^{1,2}$ 
\affil{$^1$International Centre for Radio Astronomy Research, University of Western Australia, 35 Stirling Highway, Crawley, WA 6009, Australia}
\affil{$^{2}$ARC Centre of Excellence for All Sky Astrophysics in 3 Dimensions (ASTRO 3D)}
\affil{$^3$Email: rhys.poulton@icrar.org}
}%

\jid{PASA}
\doi{10.1017/pas.\the\year.xxx}
\jyear{\the\year}

\usepackage{aas_macros}

\newcommand{\overbar}[1]{\mkern 1.5mu\overline{\mkern-1.5mu#1\mkern-1.5mu}\mkern 1.5mu}

\begin{document}

\begin{frontmatter}
\maketitle
\begin{abstract}
Merger trees harvested from cosmological $N$-body simulations encode the 
assembly histories of dark matter halos over cosmic time, and are a 
fundamental component of semi-analytical models (SAMs) of galaxy formation.
The ability to compare the tools used to construct merger trees, namely halo 
finders and tree building algorithms, in an unbiased and systematic manner 
is critical to assess the quality of merger trees. In this paper, we present 
the dendogram, a novel method to visualise merger trees, which provides a 
comprehensive characterisation of a halo's assembly history - tracking 
subhalo orbits, halo merger events, and the general evolution of halo 
properties. We show the usefulness of the dendogram as a diagnostic tool 
of merger trees by comparing halo assembly histories from a single $N$-Body 
simulation analysed with three different halo-finders -
\textsc{VELOCIraptor}, \textsc{AHF} and \textsc{Rockstar} - and 
their associated tree-builders. Based on our analysis of the resulting 
dendograms, we highlight how they have been used to motivate improvements to 
\textsc{VELOCIraptor}. The dendogram software is publicly available online, 
at: https://github.com/rhyspoulton/MergerTree-Dendograms.
\end{abstract}

\begin{keywords}
methods: numerical -- galaxies: evolution -- galaxies: halos -- dark matter
\end{keywords}
\end{frontmatter}

\section{INTRODUCTION }
\label{sec:intro}

Cosmological $N$-body simulations are a powerful and well established tool 
for studying theories of cosmic structure formation and for making  
predictions that can be compared directly to observations. By modeling the
discretised dark matter density field, these $N$-body simulations solve
for the dynamical evolution of dark matter particles under the influence of 
their mutual gravity and, in particular, track the formation and evolution 
of gravitationally bound condensations of dark matter known as halos. The
currently favoured paradigm of galaxy formation predicts that galaxies
form within these dark matter halos and that their subsequent evolution is
shaped by the growth of their host halos \citep{white_galaxy_1991}. 
Consequently, quantifying accurately how halos assemble over cosmic time is 
a fundamental requirement of cosmological $N$-body simulations. 

One of the key predictions that affect simulations of galaxy formation is 
the merging of halos and, consequently, the merging of galaxies. The timing 
of the merger and the dynamical evolution of the dark matter halo are 
important in determining galaxy properties e.g. stellar mass, galaxy colour 
and morphology \citep{boylan-kolchin_dynamical_2008}. In particular, the 
merging of halos and the galaxies contained within are intrinsically linked 
processes in Semi-Analytical Models \citep[SAMs; for review see][]{baugh_primer_2006,benson_galaxy_2010,somerville_physical_2015}. SAMs populate dark matter halos in cosmological simulations by using analytical approximations to 
self-consistently model the evolution of galaxies through cosmic time. This 
enables them to efficiently simulate large volumes of interacting galaxies 
that are in turn hosted by dark matter halos 
\citep{white_galaxy_1991,lacey_tidally_1991,cole_modeling_1991}. Because of 
the intrinsic coupling of galaxies and dark matter halos, the algorithms 
used to identify the halos and connect them across time are critically 
important. Desired characteristics are that they are robust (insensitive to 
small changes in the underlying simulation or parameter choices) and 
accurately track the halos across cosmic time through complex stages of halo 
interactions.

In essence, this means halos must be both recovered within individual 
simulation epochs, and tracked (or linked) between different simulation 
epochs. These algorithms are respectively known as halo-finders and 
tree-builders. The two most common halo-finder algorithms are the 
Spherical Overdensity (SO) method \citep{press_formation_1974} and Friends-
Of-Friends (FOF) method \citep{davis_evolution_1985}. The former groups
particles into halos by locating density peaks and growing a spherical 
volume until the mean enclosed density drops below a threshold value; the 
latter groups particles into halos by setting a linking length and 
connecting particles separated by a distance smaller than the linking length \citep{knebe_haloes_2011}.

Halos identified by halo-finders are then
connected across simulation outputs, known as snapshots, using a tree 
builder \citep[e.g.][]{lacey_merger_1993}. Tree-builders use a variety 
of properties to connect halos across snapshots, e.g. the ID of the 
particles inside the halo, halo trajectories and binding energies of the 
halos \citep[see][for a discussion on these approaches]{srisawat_sussing_2013}. Together, the combination of the
halo-finder and tree-builder allows us to build halo merger trees, which 
trace halos across snapshots and the capture merger and interactions between  neighbouring halo(s) \citep{lacey_merger_1993,roukema_merging_1997}.

Ideally, halos should be traced until they completely merge with another 
halo or are tidally disrupted. However, this is not always the case, 
ultimately coming down to the ability of the halo-finder to identify halos 
deep inside the dense environment of the host (parent) halo  
\citep[see][]{avila_sussing_2014}, and of the tree-builder to link 
them across time, even when the halo is being disrupted 
\citep[see][]{srisawat_sussing_2013}). Therefore, objective comparisons of
halo-finders and tree-builders to ensure that they produce well-behaved 
merger trees. 

Merger trees that are well-behaved do not suffer from problems such as:
\begin{description}
\item[Flip-flopping] -  The tree-builder mixes up links between two halos, but corrects it in subsequent snapshot(s) \citep{behroozi_major_2015,poole_convergence_2017}. This can lead to large change in the halo properties in the snapshots where it happens. 
\item[Branch swapping] - This is similar to flip-flopping except the tree-builder does not correct it and so the halos continue their independent evolution.
\item[Truncation] - The halo finder cannot find the halo for one or more snapshots which can lead to no good links being found so the halo is left unconnected \citep{srisawat_sussing_2013,poole_convergence_2017}.
\end{description}

To add a further complication the desired characteristics of a merger tree are at least partially subjective, and there is no theoretical approach that trivially predicts exact trees to calibrate against.


Typically, comparisons of halo-finders are done by analysing how they
perform in certain, pathological, situations, such as multiple halo 
merger events; or how well they recover global summary statistics, e.g. 
halo and subhalo mass functions \citep{knebe_haloes_2011,onions_subhaloes_2012,onions_subhaloes_2013,knebe_structure_2013,elahi_streams_2013}; or how well a particular class of major 
merger is tracked \citep{behroozi_major_2015}. Tree-builder comparisons are 
more challenging and are typically done by considering global performance 
statistics, for example, the degree of mass fluctuations present in the 
merger tree (the number of large changes in halo mass across snapshots) 
\citep{srisawat_sussing_2013}, or the length of the tree (how long halos 
have existed for in the simulation) 
\citep{avila_sussing_2014,wang_sussing_2016}. Although useful, they do not 
easily reveal what is physically happening in the simulation and, more 
importantly, what the dynamics of the systems look like. Also, it is often 
unclear what a desirable metric is, i.e. neither the longest nor shortest 
trees are likely to be the preferred outcome. More broadly, previous 
comparisons have had low information density. Orbital properties, for 
example, can be easily overlooked when examining a particular merger event 
or the overall statistics of the halos. This motivates the need for a new
analysis tool, and we focus on a novel visualisation method to capture the 
evolution of halos, their growth, interactions and tidal disruption.

\medskip

In this work, we present the merger tree dendograms, a novel visualisation 
tool that captures the evolution of a halo across the full simulation. These 
dendograms intuitively present any interactions or tidal disruption halos 
have experienced, in addition to interacting halo orbits and the distance at 
which a merger occurs. These dendograms can be used not only as a diagnostic 
tool for halo-finders and tree-builders, but also for comparison projects 
between different codes. Pathological problems are very easy to uncover, and the high information density guides the user towards potential causes and remedies. 

Using dendograms, we have been able to identify problematic events 
that caused serious artifacts in the halo-finder \textsc{VELOCIraptor} (\citealt{elahi_peaks_2011,elahi_surfs:_2018}, Ca\~nas et al., submitted, Elahi 
et al., submitted) and its corresponding tree builder \textsc{TreeFrog} (\citealt{srisawat_sussing_2013}, Elahi et al., in prep)  merger trees, which could not be detected using 
more traditional quality metrics. Their detection utilizing the dendograms have 
guided us towards significant improvements in the codes and the resultant 
merger trees which are now being used to run SAMs on our suite of N-body 
simulations \citep[SURFS][]{elahi_surfs:_2018}.

The structure of this paper is as follows. In Section 2, we specify the 
terminology that will be used. Section 3 describes the input halo catalogues 
that we use. In Section 4, we describe the merger tree dendograms, what 
information they provide and how they have been useful. For completeness, we also show merger density plots in Section 5 showing how overall statistics can highlight a problem but not suggest a solution. Finally in Section 6, we discuss and summarise the usability of these plots and highlight 
their usefulness in comparison projects.

\section{Terminology}
\label{sec:term}
\begin{figure}
\centering
\includegraphics[width=0.47\textwidth]{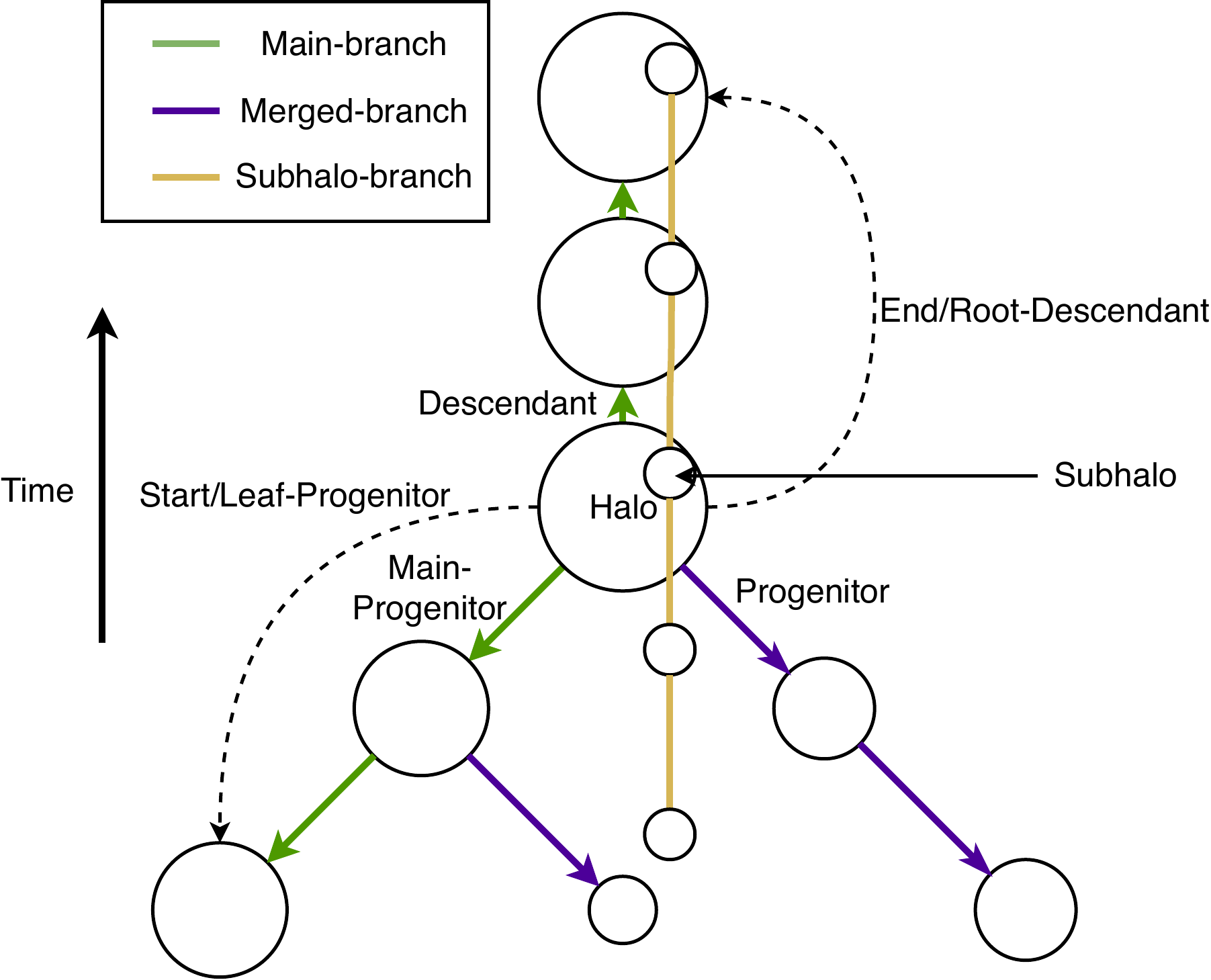}
\caption{This is a representation of the terminology used in this paper \label{fig:ETFsch}. This figure shows an example main branch, an example subhalo-branch and two merged-branches of depth 1.}
\end{figure}

For clarity, the following terminology will be used in this paper. A visual representation is shown in Figure \ref{fig:ETFsch}:
\begin{description}
\item[Halo] -  The condensation of dark matter returned by a halo-finder.
\item[Subhalo] - A halo which is not at the top of its spatial hierarchy 
(lies within one or more halos). For clarity, only subhalos of the main-branch is shown in the dendograms.
\item[Descendant] - The (sub)halo that the tree-builder has determined in 
the next snapshot(s) that the current (sub)halo goes into. Typically 
tree-builders do not allow fragmentation so a halo can only have one 
descendant.
\item[Progenitor] - The (sub)halo in the previous snapshot that a (sub)halo 
at the current snapshot has came from, as determined by the tree-builder.
\item[End/Root-Descendant] - The last/ root (sub)halo which the tree-builder has 
determined the (sub)halo at the current snapshot has ended up with in the 
simulation.
\item[Start/Leaf-Progenitor] - The first/ leaf (sub)halo which the tree-builder has 
determined the (sub)halo at the current snapshot has started with in the 
simulation.
\item[Main-progenitor] - The (sub)halo which the tree-builder has determined 
to be the main (sub)halo that the (sub)halo has come from in the previous 
snapshot and share the same Start/Leaf-Progenitor. The identification of the main-progenitor is dependent on the tree-builder used (and its configuration), therefore it can be ambiguous.
\item[Branch] -  The halos that are connected across snapshot which share a 
unique Start/Leaf-Progenitor.
\item[Main-branch] - This is the branch of the merger tree that contains the 
main progenitors for the halo in the final snapshot across cosmic history, 
used to build the merger tree. The evolutionary history of a main-branch can be dependent on the tree-builder used (and its configuration), particularly in branches with multiple mergers. These halos are connected by the green line 
in Figure \ref{fig:ETFsch}.
\item[Merged-branch] - These are branches which share the same End/Root -Descendant as the main branch but have a different Start/Leaf-Progenitor, indicating they have merged with the main branch at some point along the main branches history (travelling from Start/Leaf-Progenitor to End/Root-Descendant. These are the halos connected with purple line in Figure \ref{fig:ETFsch}.
\item[Merged-branch depth] - The depth indicates the number of times when walking a branch forward in time (from Leaf/Start-Progenitor to End/Root-Descendant) that a branch ``merges'' with another branch, i.e., instances where a halo points to a Descendant that does not point back to its Start/Leaf-Progenitor.
\item[Interacting-branch] - Branches that do not share the main branch's End/Root-Descendant but become
a subhalo of the main-branch for at least one 
snapshot. These are halos that are connected by the yellow line in Figure 
\ref{fig:ETFsch}.
\end{description}

\section{Input catalogues}

This work uses data from Synthetic UniveRses For Surveys (SURFS), a suite of 
$N$-body/ Hydrodynamical simulations \citep{elahi_surfs:_2018} spanning 
ranges of cosmological volumes to address both galaxy formation and 
cosmological surveys. The simulations assume a $\Lambda$CDM universe and use 
Planck cosmological parameters \citep{planck_collaboration_planck_2015}. All 
simulations are run with a memory-lean version of the \textsc{gadget}2 code 
with a range of box sizes from 40 to 900 Mpc/h, containing up to 10 billion 
particles. Here we focus on a small volume, moderate resolution simulation;
a 40 Mpc/h box with $512^3$ particles. For more details see 
\citep{elahi_surfs:_2018}.

The halo merger trees are built with three different halo-finders - 
\textsc{AHF} \citep{gill_evolution_2004,knollmann_ahf:_2009}, 
\textsc{Rockstar} \citep{behroozi_rockstar_2013} and \textsc{VELOCIraptor} 
(\citealt{elahi_peaks_2011,elahi_surfs:_2018}, Ca\~nas et al., submitted, Elahi 
et al., submitted) - along with their respective tree-builders -
\textsc{MergerTree}, \textsc{Consistent Trees} 
\citep{behroozi_gravitationally_2013} and \textsc{TreeFrog} 
(\citealt{srisawat_sussing_2013}, Elahi et al., in prep) - to link halos
across snapshots.

We test two types of halo-finders - configuration space and phase space 
halo-finders. Configuration space halos-finders generally use either 
density or position information to find halos; this enables them to easily
pick out halos, but they have difficulty identifying sub-structure. Phase
space halo-finders use both position and velocity information to identify 
sub-structure; the extra information enables a much more robust 
identification of subhalos.

All of the tree-builders used here are particle correlators, which use particles bound to 
halos to link them across snapshots, effectively identifiying all possible connections for every halo. They then maximize a merit function(s) to 
find a single descendant and main progenitor \citep[for further details see][]{srisawat_sussing_2013}. Both \textsc{Consistent Trees} and 
\textsc{TreeFrog} have corrections to account for missing halos, discussed 
below.

\subsection{\textsc{AHF} \& \textsc{MergerTree}}

\textsc{AHF} first finds local overdensities in an adaptively smoothed 
density field to find possible halo centers. Then particles that are 
gravitationally bound to density peaks are identified to construct the 
halo \citep{knollmann_ahf:_2009}. \textsc{MergerTree} is a particle 
correlator, as discussed above.   \textsc{MergerTree} only identifies connections one snapshot in the past and by default will give a graph, but a merit can be used to create a mergertree.

\subsection{{\sc \bf Rockstar} \& \textsc{Consistent Trees}}

\textsc{Rockstar} is a phase-space halo-finder that uses an adaptive 
hierarchical refinement of six Dimensional Friends-Of-Friends (6DFOF) and 
one time dimension, which enables explicit tracking of merged structure. 
Its tree-builder, \textsc{Consistent Trees} corrects for missing halos 
by using halo trajectories from gravitationally evolving positions and 
velocities of halos between time-steps. By making use of information from 
surrounding snapshots, it can correct for missing or extraneous halos 
\citep{behroozi_gravitationally_2013}. This is required in the case when 
\textsc{Rockstar} can no longer find the subhalo because it is too close to 
its host's centre, but it is not fully merged so correction is needed 
to allow the subhalo to merge.
 
 \subsection{  \textsc{\bf VELOCIraptor} \& \textsc{\bf TreeFrog}}
 
\textsc{VELOCIraptor} is a 6DFOF halo-finder that first uses the 3DFOF 
algorithm to identify halos and then identifies substructure using the 
full 6DFOF information to robustly identify not only subhalos but also 
tidal streams from disrupted halos \citep{elahi_streams_2013}. 
Its tree-builder \textsc{TreeFrog} has the additional capability to link
halos over more than one snapshot, which enables a halo to be linked even if 
it is not found by the halo finder in one or more snapshots away  
(\citealt{srisawat_sussing_2013}, Elahi et al., in prep) . After testing, 
the linking is set to connect up halos over four snapshots in both the old and 
new \textsc{VELOCIraptor} \& \textsc{TreeFrog} catalogues. \\

\begin{figure}[ht]
\centering
 \includegraphics[width=0.47\textwidth]{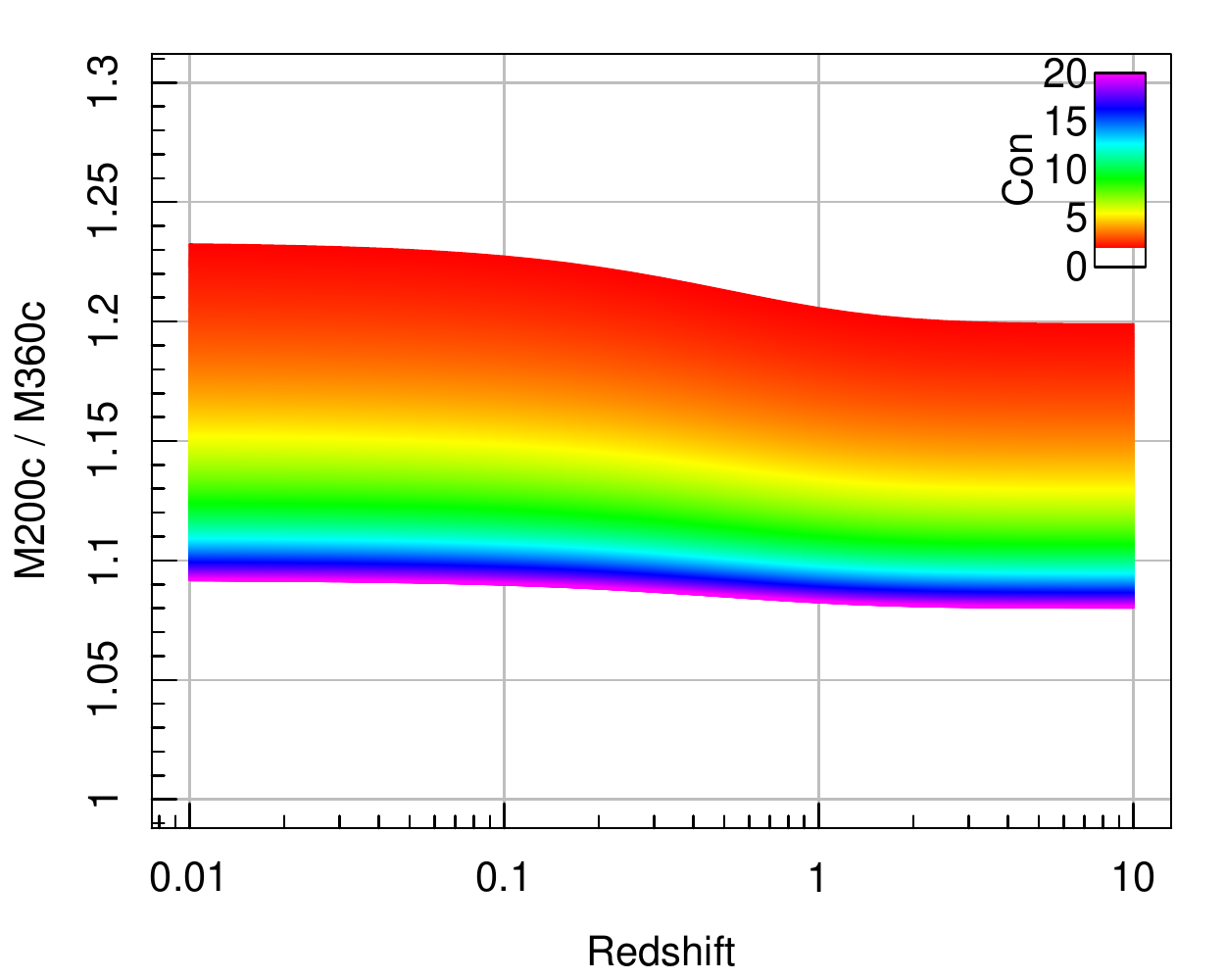}
 \caption{Visual representation of how the different definitions of M$_{\rm vir}$ changes as a function of redshift and the halos concentration.}
  \label{fig:mvirDiff} 
\end{figure}
 \subsection{ {Mass Definitions}}
Different halo-finders use different definitions for virial mass 
(M$_{\textrm{vir}}$). \textsc{AHF} uses M$_{200,\textrm{crit}}$(mass 
contained within the region with  $\overbar{\rho}$ = 
200$\rho_{\mathrm{crit}})$; \textsc{VELOCIraptor} uses all of the 
definitions of M$_{\rm vir}$, so we select M$_{200,\textrm{crit}}$; 
\textsc{Rockstar} uses the definition from \citep{bryan_statistical_1998}, 
which corresponds to M$_{\rm 360,crit}$ (mass contained within the region 
with  $\overbar{\rho}$ = 360$\rho_{\mathrm{crit}}$) times the background 
density at z = 0. A visual representation of the differences as they change with halos concentration is shown in Figure \ref{fig:mvirDiff}, which demonstrates that the two definitions vary by only 10-20\% depending on the concentration.  All masses in this paper will be inclusive, so they
include the mass of any subhalo that lies within the halo's virial radius 
(R$_{\mathrm{vir}}$, the radius containing M$_{\textrm{vir}}$ ); see 
appendix \ref{app:exclMass} for an example of a merger tree with exclusive 
masses.

\section{Merger Tree Dendograms}

\begin{figure}[ht]
\centering
 \includegraphics[width=0.3\textwidth]{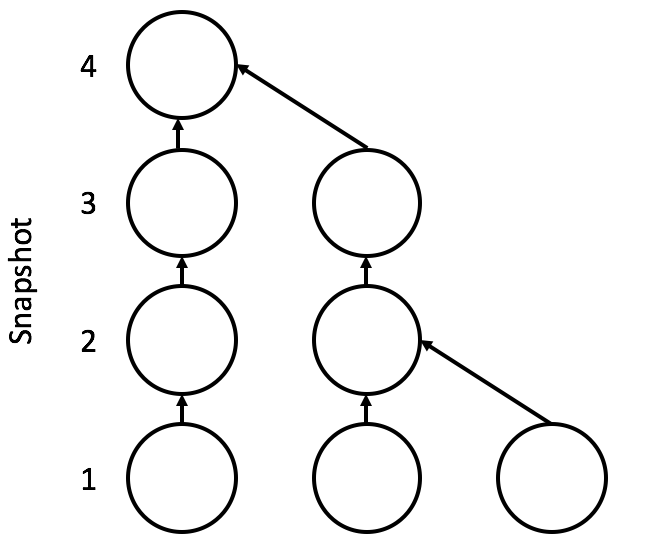}
 \caption{An example of how merger trees are typically represented. This plot shows the merger history for a single halo in the final snapshot, showing the main-branch (the halos directly beneath it) and the merging branches. However, it does not give a insight of how the halos are interacting or how far away a merger happens.}
  \label{fig:SimpleMT} 
\end{figure}

The traditional way of showing merger trees is shown in Figure 
\ref{fig:SimpleMT} \citep[see][]{roukema_merging_1997}. This diagram only 
shows the rough merger history of a single halo in the final snapshot and 
does not convey the mass evolution, a critical piece of information. A 
common improvement is the addition of the halo masses in the tree \cite[see][for examples]{de_lucia_hierarchical_2007,tweed_building_2009,hirschmann_stellar_2015,mcalpine_eagle_2016,naab_theoretical_2017}. However, it is not possible to 
extract further information, e.g. the separation the merger occurred at, or the orbit of the merging halo. 

\begin{figure*}[ht]
\centering
 \includegraphics[width=0.9\textwidth]{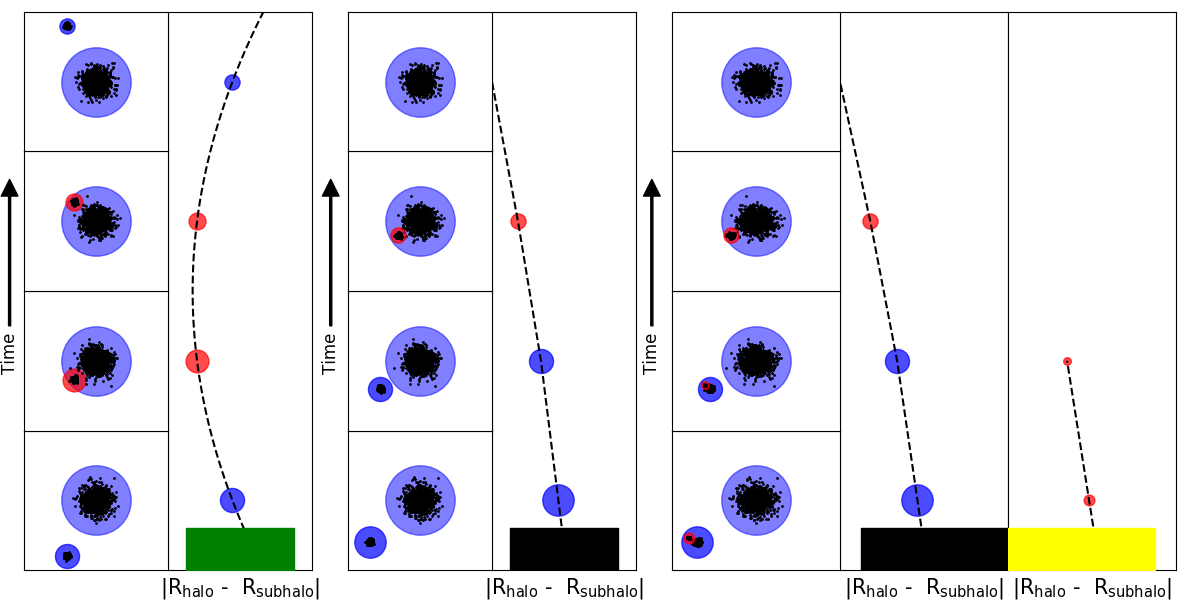}
 \caption{This diagram shows how the right sub-panels condense the orbits and 
 interactions of halos presented in the left sub-panels. The black points are 
 the halo particles, while the blue and red circles correspond to halos and 
 subhalos respectively, with the size of the circle representing the mass of 
 the (sub)halo at each time-step. The dashed 
 black line shows a quadratic (left panel) and linear (middle and right panel) splines of the halo position, demonstrating the path that the halo would most likely take. Here we show from left to right: a fly-by (shown by the green 
 block); single merger event (shown by the black block); and multi-merger 
 event with a small (merged-branch of depth 2) halo merging with a larger halo (shown 
 by the yellow block connected with the black block). 
 \label{fig:SimpDendo}}
\end{figure*}

For this reason, we propose the Merger Tree Dendogram, which contains the 
entire merger history of a single halo, including any halo it has ever 
interacted with across cosmic history. The plot contains information on:
\begin{itemize}
	\item \textbf{Interaction \& merger history} -  The standard merger 
    history of a single halo in the final snapshot (known as the main 
    branch), and other halos it merged or interacted with and their 
    respective merger history (known as merged-branches).
	\item \textbf{Evolution of mass (or other quantities such as V$_{\rm max}$)} 
    - The mass evolution of the main and merged-branches throughout the 
    simulation.
	\item \textbf{Orbits} - The orbits of merged-branches and subhalos 
    around the main branch.
	\item \textbf{Lifetimes} - How long a (sub)halo exists before it merges.
	\item \textbf{Radial distance} - How far away the merged-branches merge 
    with the main branch.
\end{itemize}

A pedagogical example of the information content is presented in Figure 
\ref{fig:SimpDendo}, which shows how the main branch can interact with 
its merged-/subhalo branches. The left sub-panel of each panel shows a 
projection of the halos in the simulation, while the right sub-panel 
shows the corresponding dendogram. We show different scenarios from a flyby 
(leftmost) to a merger event (rightmost). This shows how the dendograms can
simplify 3D orbits into a 1D plot.

An example dendogram tracing the full interaction history of a halo from a 
full cosmological simulation is shown in Figure \ref{fig:oldVELgoodMT}. Its 
history is reconstructed using an older version of 
\textsc{VELOCIraptor}+\textsc{TreeFrog} halo catalogue from the 40 Mpc/$h$, 
512$^3$ particles SURFS simulation \citep{elahi_surfs:_2018}. 

The dendogram traces the full interaction and merger history of a single halo in the final snapshot. This plot only shows a merged-branch depth of 1, as displayed by the “Merged branch depth” indicator. The left-most panel shows the evolution of the main branch and the panels on the right show interacting or merged branches. The size of the line by default represents M$_{\rm vir}$ at each snapshot and the colour represents the type of halo: blue for a halo at the top of the spatial hierarchy (i.e. the central halo), or red for a subhalo. The y-axis shows the snapshot number in the simulation. For the left-most panel, the x-axis shows the Euclidean co-moving distance that the main branch halo has moved in the simulation and for the rest of the panels on the right, the x-axis shows the ratio of the radial distance to the main branch (R$_{\rm orbit}$) to the virial radius of the main branch (R$_{\rm vir,parent}$). The colour of the bar at the bottom indicates if it is a merged-branch (black), if it has sub-merged-branch (yellow), or a  subhalo branch (green). The dashed line in the panels represents one R$_{\rm vir,parent}$ and the panels are plotted up to 2.5 R$_{\rm vir,parent}$. The number on top of each of the panels is the maximum M$_{\rm vir}$ within each branch in 10$^{10}$ M$_{\odot}$. Many intuitive characteristics jump out from this high information density plot. E.g.\ as you might expect, most halos (blue) become subhalos (red) very close to the virial radius of the parent halo they merge into (dashed line). Also, the parent halo (left blue) increases in mass fairly monotonically, whilst subhalos (red) decrease in mass fairly monotonically due to dynamical friction within the R$_{\rm vir,parent}$.

The dendogram code is written in Python 3 and is able to read a variety of 
halo merger tree catalogue formats. The code (appendix \ref{app:code}) 
requires that trees are in Efficient Tree Format (ETF), which is a 
reduced version of the SUSSING MERGER TREE format 
\citep{thomas_sussing_2015}, and so a conversion tool has to be built to 
convert the data into this format\footnote{Please email 
rhys.poulton@icrar.org for assistance in building a conversion tool if 
needed}. Conversion tools already exists for \textsc{VELOCIraptor} + 
\textsc{TreeFrog} \citep{elahi_streams_2013,elahi_peaks_2011}; 
\textsc{AHF} + \textsc{MergerTree} \citep{knollmann_ahf:_2009}; Millennium 
\citep{springel_simulations_2005}; and \textsc{Rockstar} + 
\textsc{Consistent Trees} 
\citep{behroozi_rockstar_2013,behroozi_gravitationally_2013}. The 
code also accepts the SUSSING MERGER TREE format 
\citep{thomas_sussing_2015}. Details of the code are in appendix 
\ref{app:code} and the ETF is in \ref{app:form}.

\subsection{As a diagnostic tool} \label{sec:diag}

\begin{figure*}[hp]
\centering
 \includegraphics[width=\textwidth]{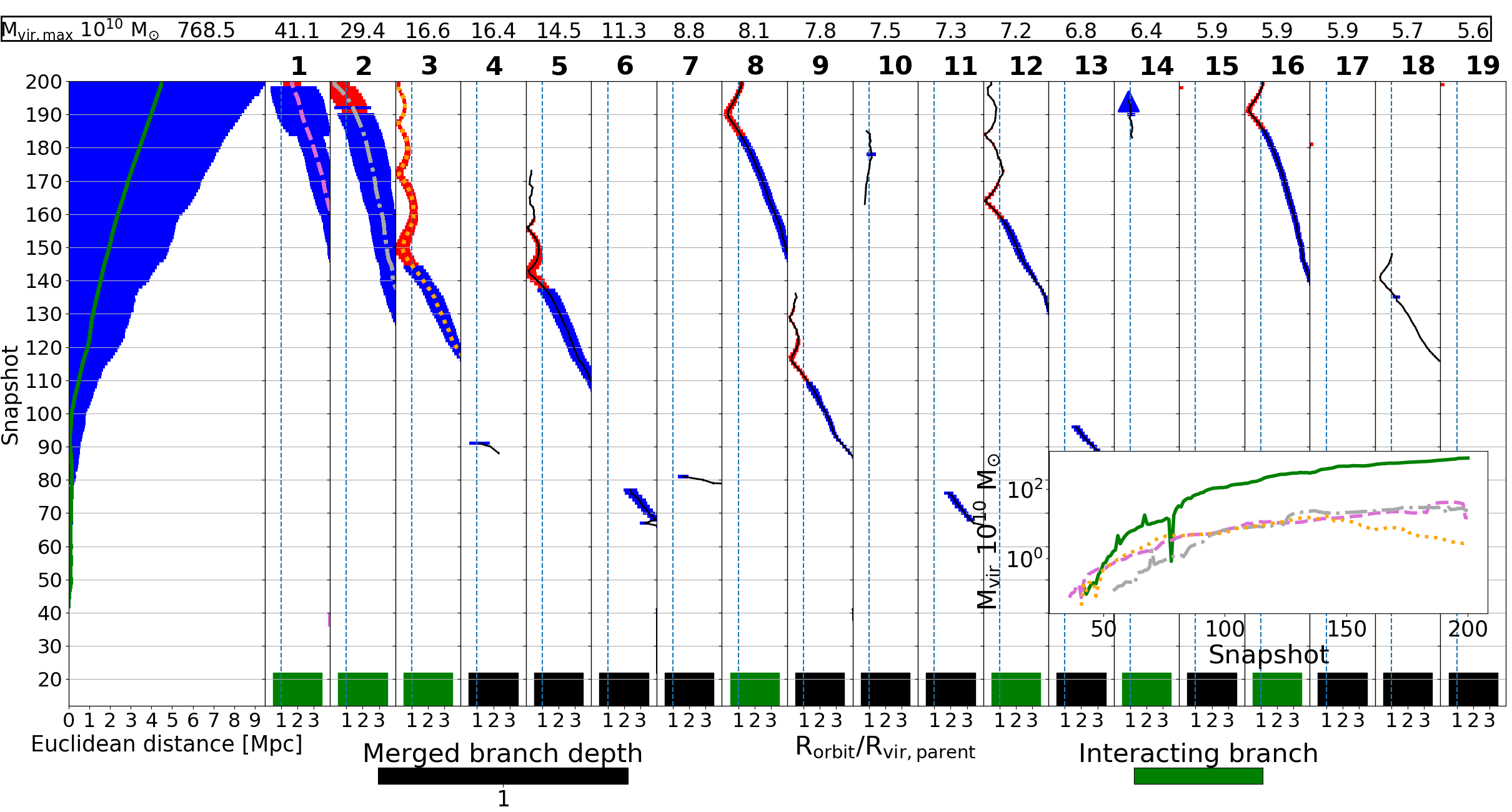}
 \caption{An example merger tree dendogram from the the old \textsc{VELOCIraptor} + \textsc{TreeFrog} catalogue, here on referred to as HALO1. The Euclidean distance shown for the main panel shows the comoving distance it has travelled with reference to its formation position. The inset plot in the Figure shows the mass history for the four largest branches, which are represented by the different coloured lines/ line styles in the branch. The blue and red points correspond to halos and subhalos respectively. The triangle shows that this branch has become a subhalo of the branch of interest for at least one snapshot, but then has merged with another branch. In this case it has merged with branch 2 shown in this plot. }
 \label{fig:oldVELgoodMT}
\end{figure*}

\begin{figure*}[hp]
\centering
 \includegraphics[width=\textwidth]{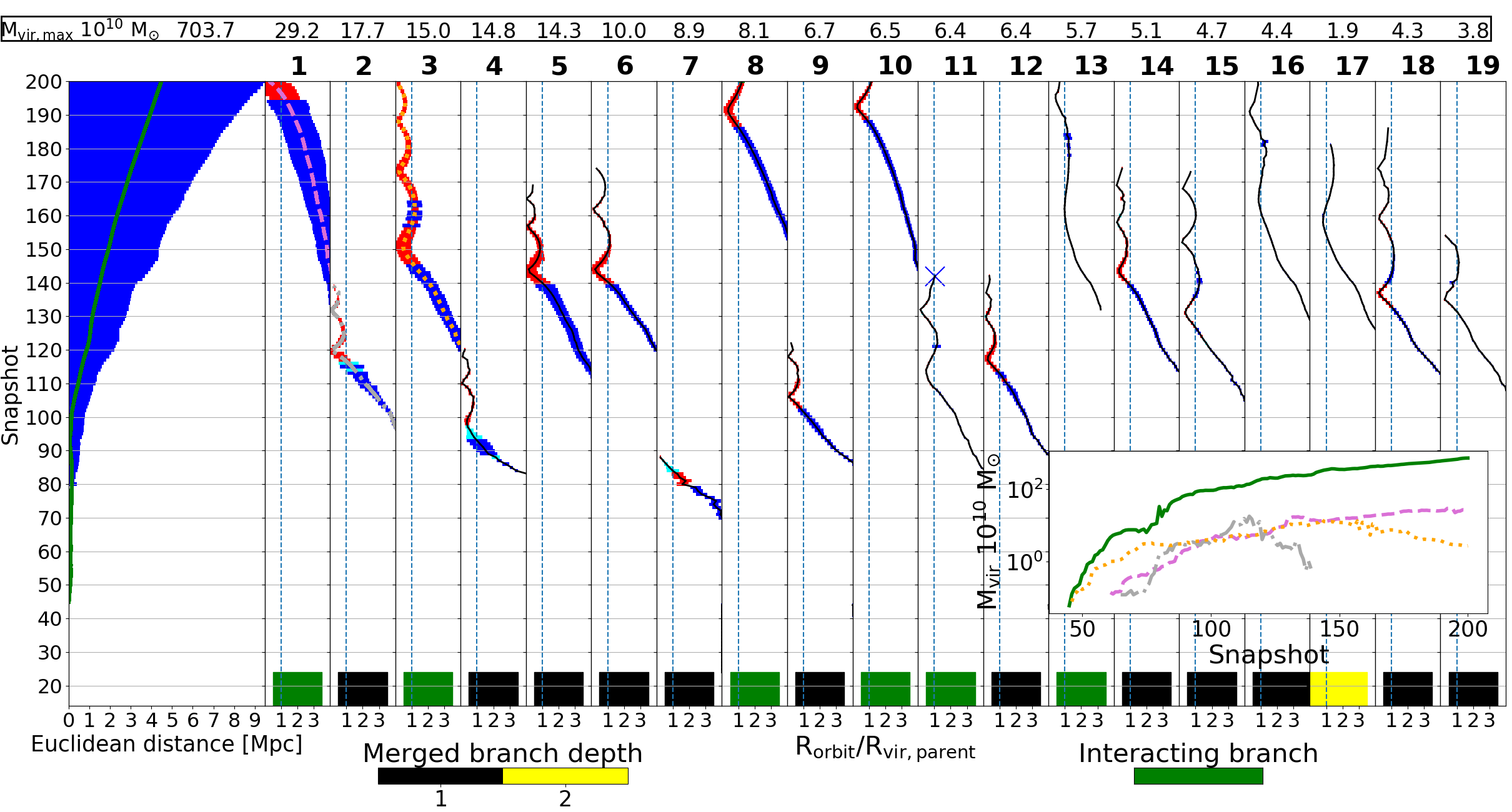}
 \caption{HALO1 merger tree reconstructed by the updated \textsc{VELOCIraptor} + \textsc{TreeFrog} + \textsc{WhereWolf} catalogue. The cyan points are the \textsc{WhereWolf} halos that are inserted into the catalogue. For clarity, the branches with a merged-branch depth of 2 are only shown if they exist  80 snapshots before they merge with a merged-branch depth of 1. }
 \label{fig:VELgoodMT}
\end{figure*}

The dendogram allows for a visual inspection of the orbital and mass accretion histories of halos, providing an insight into the interaction history of  the main branch with its merged-/subhalo-branches. It can be used to identify the problems occurring due to the particular code used, in addition to the frequency of the problems occurring. It has been useful in identifying problems with the old version of \textsc{VELOCIraptor} and \textsc{TreeFrog} which include:
\begin{enumerate}
\item The merging of halos with their host well outside of the host R$_{\mathrm{vir,parent}}$.
\item The identification of the incorrect main-progenitor/ descendant, causing an abrupt change in the M$_{\mathrm{vir}}$ of the halo.
\item The halo not being identified by \textsc{VELOCIraptor}, leaving a break in the existence of the halo.
\end{enumerate}

The identified problems can be seen in the dendogram in Figure \ref{fig:oldVELgoodMT}. The key problem is that some merged-branches do not become a subhalo before merging since they are well outside of their R$_{\mathrm{vir,parent}}$ (see branches 4,6,7,11,13), and so they suffer from the first and third problem identified above. This is most likely an issue with either \textsc{VELOCIraptor} being unable to identify the halo, or \textsc{TreeFrog} not connecting the halos up across snapshot. The "over-merging" problem occurs during some major mergers but it is not immediately evident if one focuses on the statistical median and scatter of the halo occupation.

Furthermore from the inset plot in Figure \ref{fig:oldVELgoodMT}, the main branch (green line) shows a large change in mass in snapshot 80. This large change in mass occurs because of a merger of comparable masses that happens at that snapshot, which makes it difficult to reconstruct the halo's mass. This can be traced to the old version of \textsc{VELOCIraptor} struggling in such cases. Moreover there are halos that have a short existence in branches 10, 14, 15, 17 and 19, indicating that either \textsc{VELOCIraptor} cannot accurately reconstruct these halos histories or  \textsc{TreeFrog} is having issues in connecting up these halos. 

Upon identification of these problems, modifications were made to the \textsc{VELOCIraptor} and \textsc{TreeFrog} algorithms. These changes have improved the halo tracking in merger events, improving the reconstructed orbital evolution of subhalos. These improvements also enable a detailed study on the dynamical friction timescales for the merging halo, which will be left for a future paper (Poulton et al., in prep). The detailed discussion of the optimizations/ improvements have been left for a future paper (Elahi et al., in prep).

Figure \ref{fig:VELgoodMT} shows an example of the same merger tree as shown in Figure \ref{fig:oldVELgoodMT} after the optimizations have been implemented. Panels in Figure \ref{fig:VELgoodMT} have a different ordering to those in Figure \ref{fig:oldVELgoodMT}, which is due to \textsc{VELOCIraptor}'s improved ability to pick out substructure in the updated version and its subhalo classification. This leads to reduced masses being assigned to the main branch in Figure \ref{fig:VELgoodMT} because it is not assigning the subhalo's mass to the host's overdense region. 

The cyan lines in Figure \ref{fig:VELgoodMT} show the corrections made based on a code under development known as \textsc{WhereWolf} (Poulton et al., in prep). This code was developed because \textsc{TreeFrog}'s ability to link up halos across snapshots is limited, even with optimised merit schemes, and this can leave gaps in halo evolution. \textsc{WhereWolf} fills those gaps by tracking halo particles from the snapshot in which the halo was lost. \textsc{WhereWolf} performs a bound calculation and uses the most bound particles to define the position the next snapshot; it also tracks a halo until it is dispersed or until a match to a \textsc{VELOCIraptor} halo that was previously unlinked is found (shown in the  fourth  and seventh  branches on Figure \ref{fig:VELgoodMT}). This enables a further tracking of the subhalos even when \textsc{VELOCIraptor} can no longer find them. A detailed description of the code will be available in Elahi et al., (in prep) and Poulton et al., (in prep). 

Compared to the inset plot in Figure \ref{fig:VELgoodMT}, the main branch has a much smoother mass accretion history in the inset plot in Figure \ref{fig:oldVELgoodMT}. The large mass loss in Figure \ref{fig:oldVELgoodMT} at snapshot 80 has become an increase in mass in Figure \ref{fig:VELgoodMT}. This happens because the updated \textsc{VELOCIraptor} can recover both of the halos and also includes the mass of other merging subhalos not present on this plot.

These dendograms show that most of the problems listed above have been 
addressed. However in Figure \ref{fig:VELgoodMT}, the 13$^{\rm th}$, 16$^{\rm th}$ and 19$^{\rm th}$ branches undergo a large change in mass. This is a result of the halo changing from a subhalo to halo. A subhalo mass is exclusive, but as a halo its mass is inclusive (includes the mass of its own subhalos). This is the same case in Figure \ref{fig:oldVELgoodMT} in the 10$^{\rm th}$ and 18$^{\rm th}$ branches where it undergoes a large change in mass and does not affect the mass exclusive to the halo.

\subsection{As a comparison tool}

\begin{figure*}[hp]
\centering
 \includegraphics[width=\textwidth]{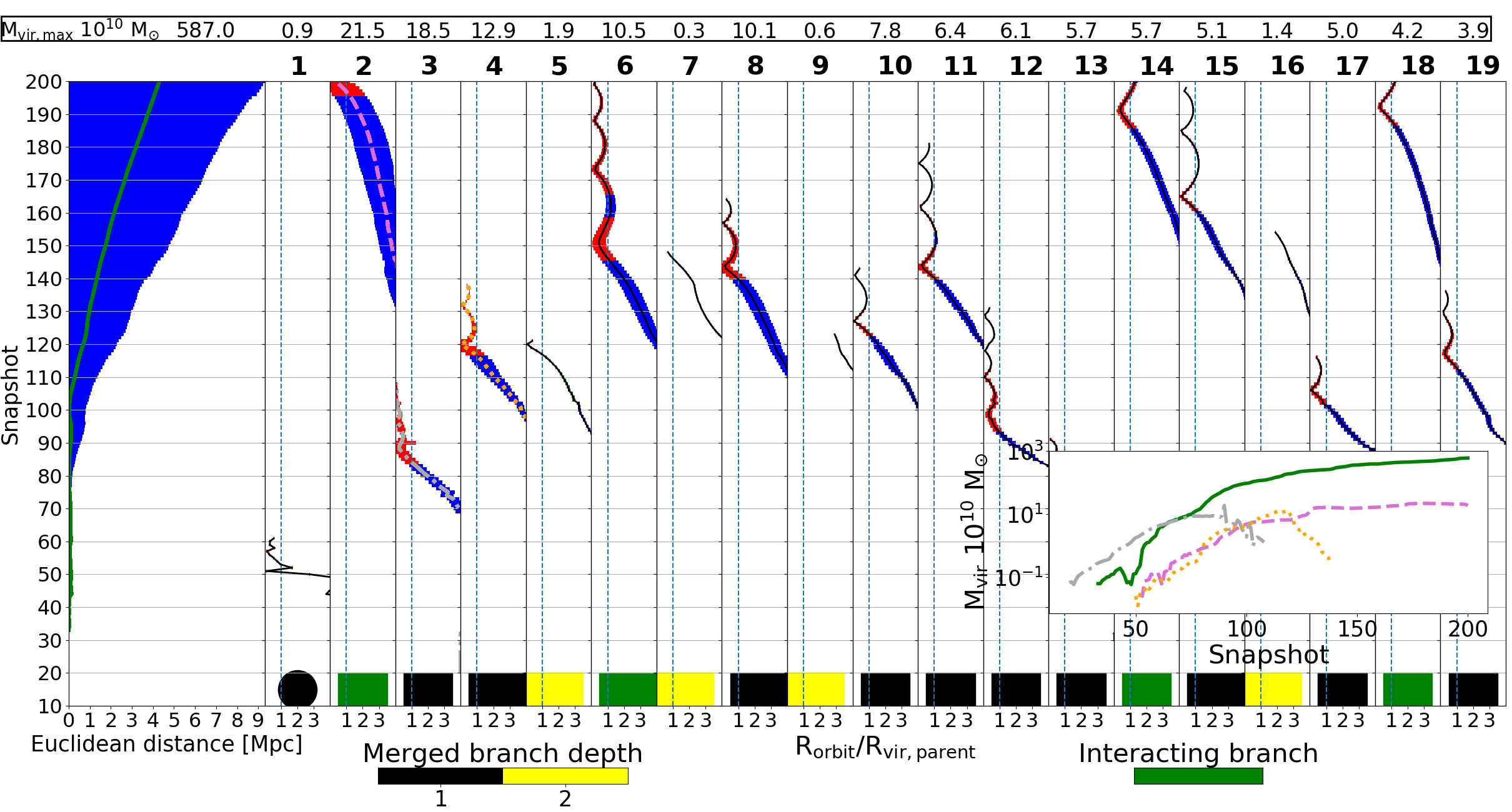}
 \caption{HALO1 merger tree reconstructed by \textsc{Rockstar} + \textsc{Consistent Trees}. The circle at the bottom of the first merged-branch shows that this branch temporally hosted the main branch for at least one snapshot}
 \label{fig:RockgoodMT}
\end{figure*}

  \begin{figure*}[hp]
\centering
 \includegraphics[width=\textwidth]{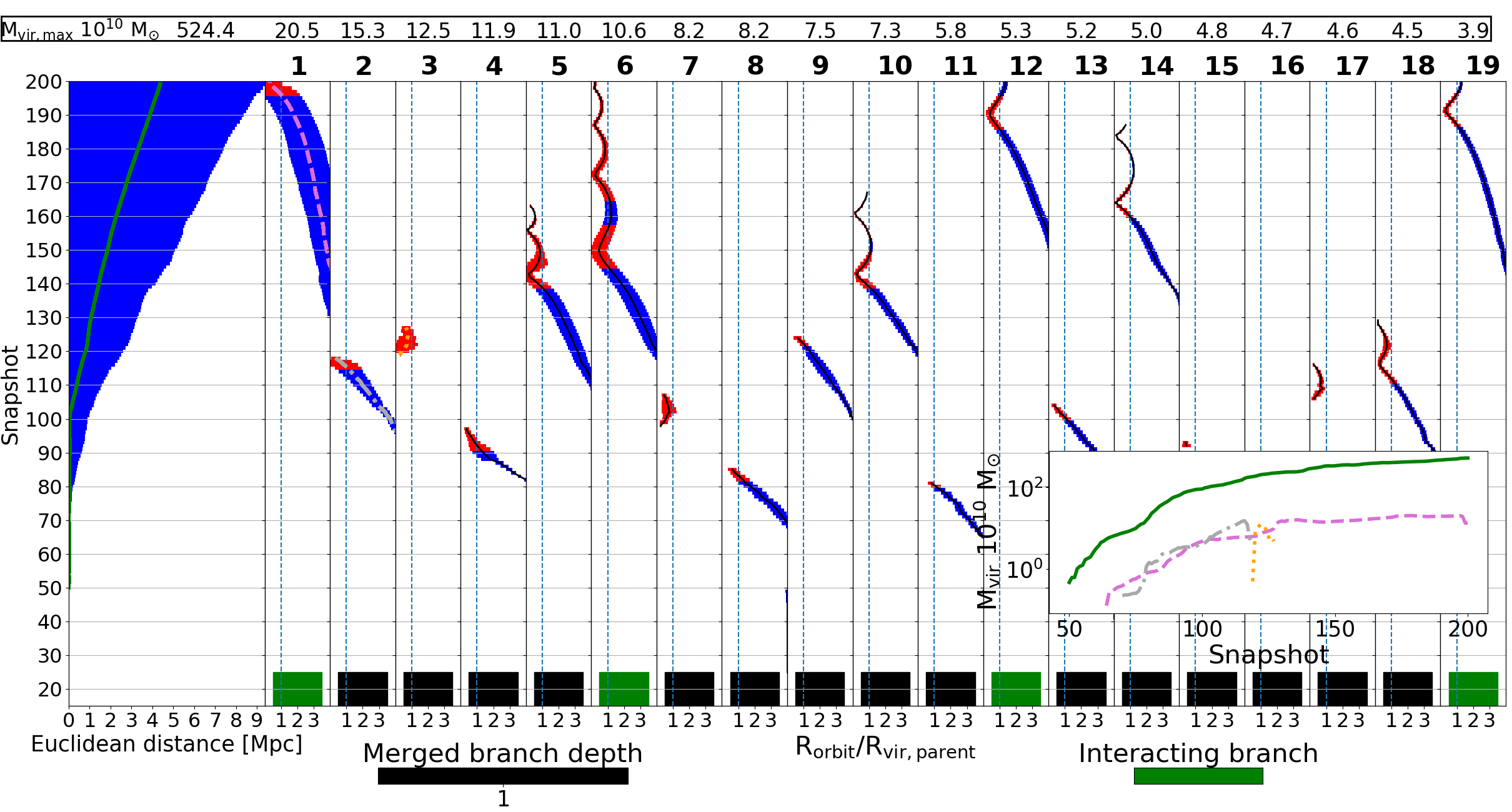}
 \caption{HALO1 merger tree reconstructed by \textsc{AHF} +\textsc{MergerTree}.}
  \label{fig:AHFgoodMT}
\end{figure*}

\subsubsection{Tree with a simple merger history}

To compare different code families, we create dendograms using both \textsc{AHF} and \textsc{Rockstar} 
merger trees, following the same halo shown in Figures \ref{fig:VELgoodMT} 
and \ref{fig:oldVELgoodMT} so that a direct comparison can be made with 
\textsc{VELOCIraptor} merger tree.

An example of a \textsc{Rockstar} merger tree is shown in Figure \ref{fig:RockgoodMT}. 
\textsc{Rockstar} does a good job of tracking that halo,showing similar 
halo evolution to the one seen in results from the updated 
\textsc{VELOCIraptor} + \textsc{TreeFrog} dendogram. In snapshot 49, the main branch 
is temporarily hosted by a merged-branch, as indicated by the circle at the 
bottom of the panel. This occurs because as the main-branch 
undergoes a rapid loss in mass \textsc{Rockstar} cannot reconstruct the 
mass of the halo for this snapshot because of the similar mass 
merger happening with few particles. The main-branch halo grows in mass again in snapshots 
50 and 51, causing the merged-branch to be well within 
R$_{\rm vir,parent}$ at snapshot 51.
Note that the third branch, at 
snapshot 90 also undergoes a large growth in mass which is lost in the next 
snapshot. The inset plot of Figure \ref{fig:RockgoodMT}  shows this 
sudden growth. This happens just as the halo is moving outward, indicating 
that it has included some of the mass of the main branch's halo.

The \textsc{AHF} + \textsc{MergerTree} dendogram in Figure 
\ref{fig:AHFgoodMT} shows a smoother mass evolution of the main branch than 
the dendograms produced by \textsc{VELOCIraptor} and \textsc{Rockstar}. 
However, subhalos shrink as they move inward and then grow again as they
move outward, a known artifact of configuration space halo-finders
\citep{muldrew_measures_2012}. A further issue shown in this dendogram is 
that branches 3, 7, 15 and 17 have no connection to infalling halos and first 
appear inside the halo. These seem to be halos from branches 2, 4, 8 and 13 
respectively that have gone through pericentric passage; \textsc{AHF}, 
being a configuration space halo-finder, can no longer pick them out from 
the background distribution and so they are lost for a few snapshots. 
Because \textsc{MergerTree} can only link halos one snapshot apart, (sub)halos 
that are not identified by the halo-finder for more than two snapshots will 
not be linked and so two separate branches are created. This can lead to 
problems within SAMs; the halo will be assumed to have merged --- and hence, so do
the galaxies hosted by the halo with the timing of the galaxy merger dependent on the SAM --- even though the halo 
have yet to merge with the host halo. The SAM would then
incorrectly assume that another halo has formed, which it may populate 
with a new galaxy, and this can have a large effect on the central galaxy 
it was going to merge with. This is a well known short-coming of 
configuration space halo finders  
\citep[see][]{knebe_haloes_2011,onions_subhaloes_2012,knebe_structure_2013,cautun_subhalo_2014,poole_convergence_2017,elahi_surfs:_2018}. 

\subsubsection{Tree with a violent merger history}

We now test how different halo-finders + tree-builders perform in the 
challenging case of a quadruple merger event, which we refer as QUAD1.
This is shown in Figure \ref{fig:3dplot}, where we see several halos with
similar masses merge in a short period of time. During this event, 
it is difficult, both objectively or subjectively, to determine the 
particles belonging to each halo.

\begin{center}
\captionsetup{type=figure}
 \includegraphics[width=0.48\textwidth]{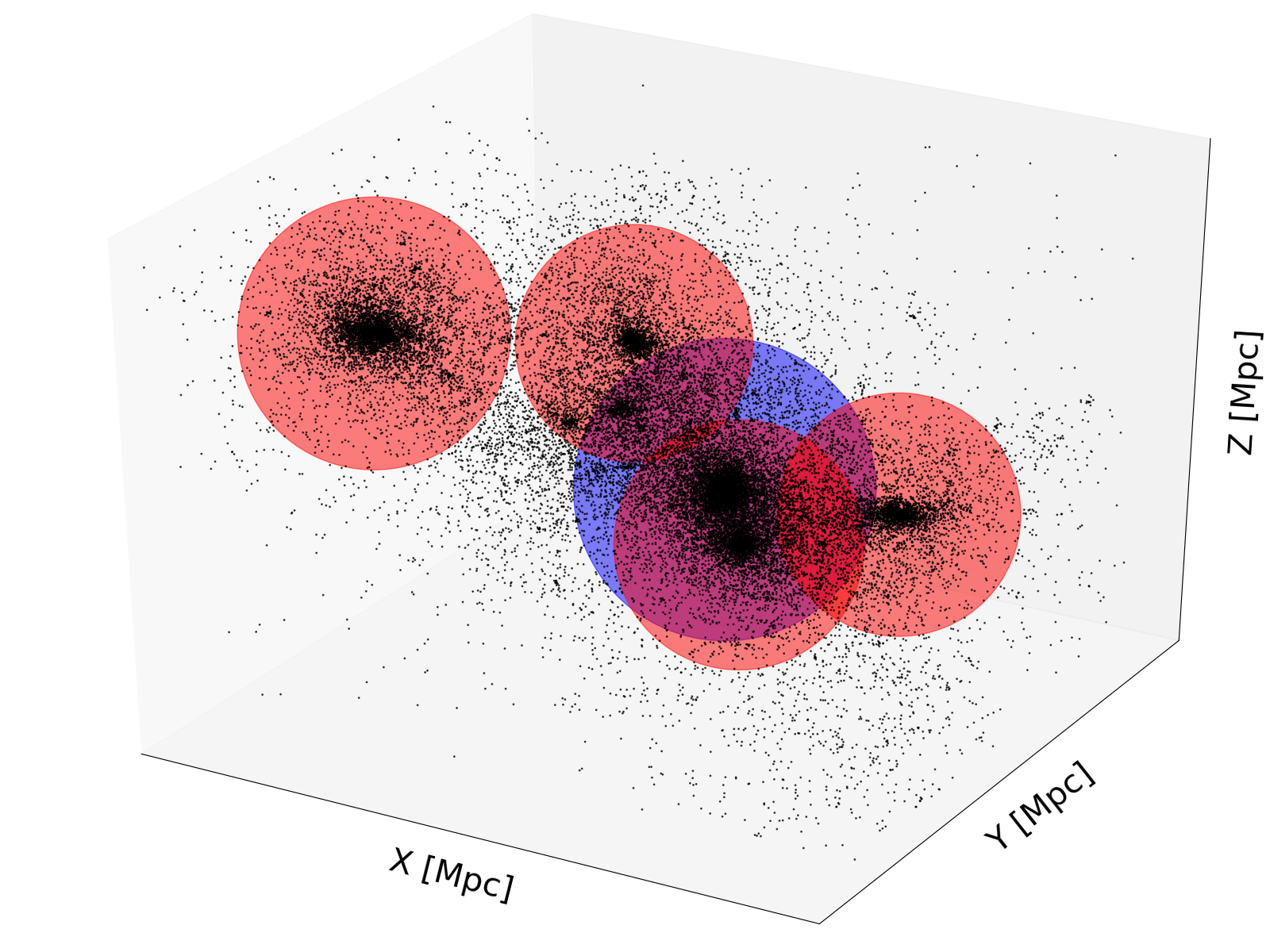}
 \captionof{figure}{This 3D plot shows the event of a quadruple merger (QUAD1). Only the five largest halos from \textsc{VELOCIraptor} are shown in this plot at snapshot 190 in the simulation. The blue halo represents the halo from main branch and the red ones are the halos from the subhalo-branches associated with the main branch.  \label{fig:3dplot}}
\end{center}

\begin{figure*}[hp]
\centering
 \includegraphics[width=\textwidth]{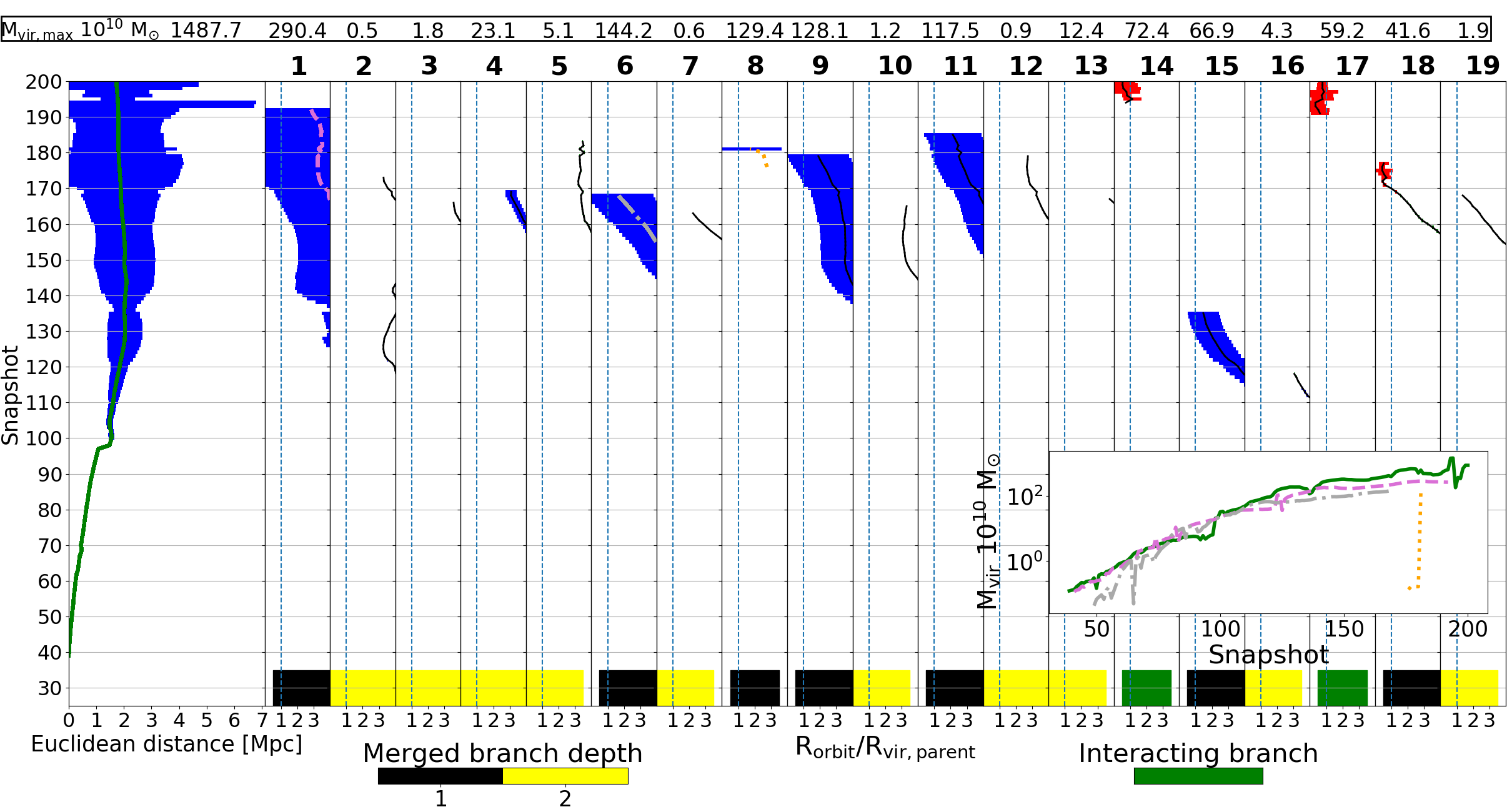}
 \caption{QUAD1 merger tree from the old \textsc{VELOCIraptor} + \textsc{TreeFrog} catalogue.}
 \label{fig:VELoldDendoBad}
\end{figure*}

\begin{figure*}[hp]
\centering 
\includegraphics[width=\textwidth]{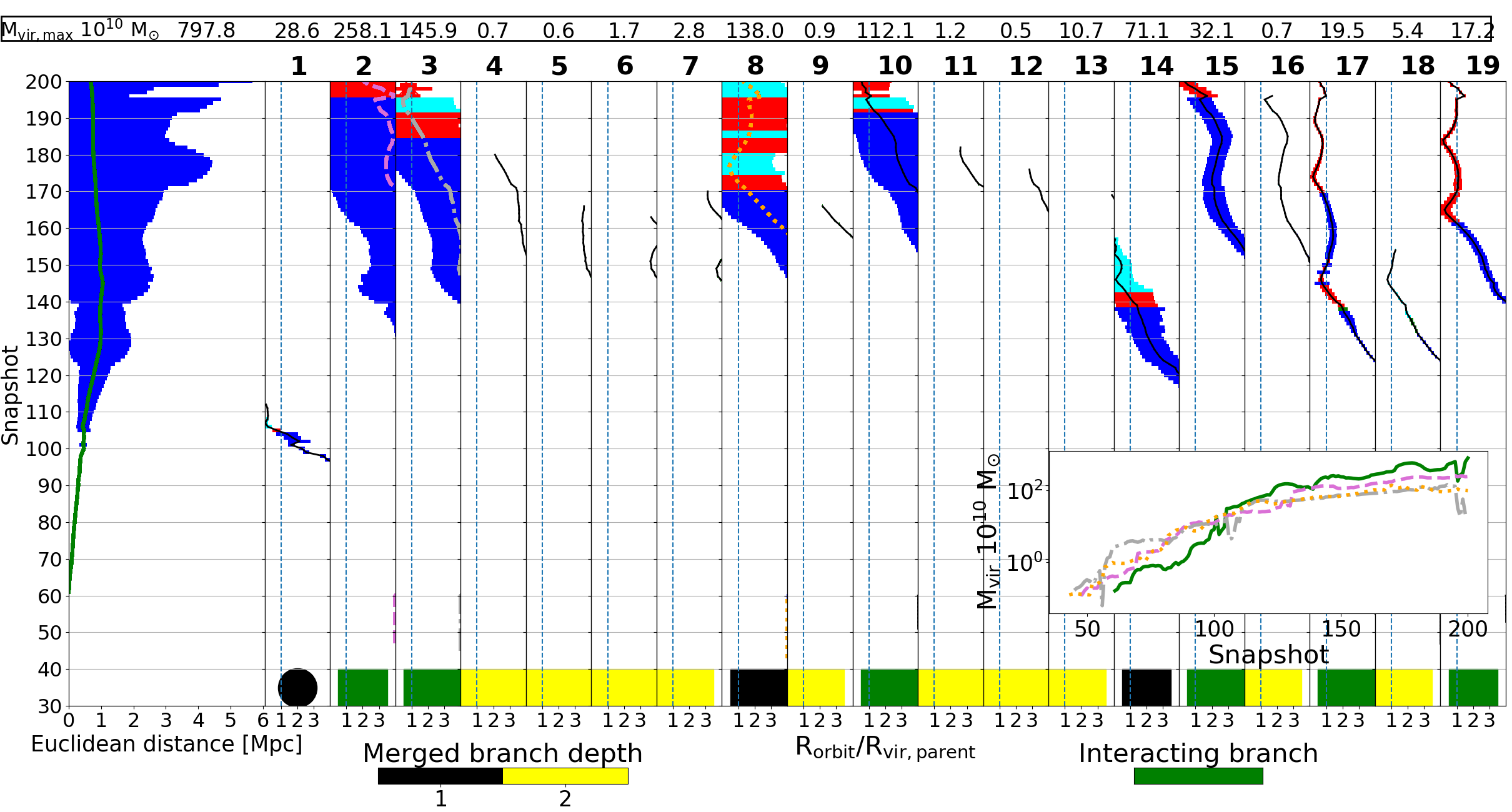}
 \caption{QUAD1 merger tree from the updated \textsc{VELOCIraptor} + \textsc{TreeFrog} + \textsc{WhereWolf} catalogue.}
 \label{fig:VELDendoBad}
\end{figure*}

The dendogram for QUAD1 constructed using the older version of 
\textsc{VELOCIraptor} is shown in Figure \ref{fig:VELoldDendoBad}. Here 
it is clear that the old \textsc{VELOCIraptor} has problems tracking 
merging halos because they never become subhalos, instead disappearing (i.e. 
merging) well outside R$_{\rm vir, parent}$. The early merging of halos is 
the same problem identified in Figure \ref{fig:oldVELgoodMT}, indicating 
that it is a recurring problem with the old \textsc{VELOCIraptor}. 
When early merging happens, the main branch does not have the mass of the 
"merged" branch associated with it because the merger happens outside 
R$_{\rm vir,parent}$. The main branch in Figure \ref{fig:VELoldDendoBad} 
also undergoes a large change in mass from snapshots 192 to 200, which is caused by \textsc{VELOCIraptor} associating the mass of the merging branches with itself. This seems to be due to the halos lying within the same overdensity envelope. However, a few snapshots later this connection is broken which causes the rapid decrease in mass. Branch 18 also undergoes a large increase in mass, suggesting that these halos may be the incorrectly connected halos from branch six.  In addition, branches 14 and 17 have no connection to infalling halos, which is the same in \ref{fig:oldVELgoodMT}, further indicating that there is a recurring problem with \textsc{TreeFrog} in connecting up the halo correctly. These seem to be the halos from branches 1 and 11 respectively. Finally, the eighth branch is connected up to the wrong halo - as shown by the large change in its mass as shown in the inset plot - demonstrating that this is a problem caused by \textsc{TreeFrog} identifying the incorrect progenitor/descendant, which is the second problem discussed in Section \ref{sec:diag}. 

For comparison, the updated \textsc{VELOCIraptor} + \textsc{TreeFrog} along 
with \textsc{WhereWolf} is shown in Figure \ref{fig:VELDendoBad}. It can be 
seen that the main branch fluctuates in mass. The fluctuation is caused by 
the mass being inclusive and the halo's mass swaps from being a halo to being a 
subhalo (See appendix \ref{app:exclMass} for the dendogram with just 
exclusive masses and also before \textsc{WhereWolf} has been run on the 
catalogue). The mass fluctuation happens just as the 
third branch passes within R$_{\rm vir,parent}$, and so its mass is 
associated with the main branch. Similarly, the large decrease of mass at 
snapshot 195 happens just as the third branch moves out of 
R$_{\rm vir,parent}$, and so its mass is not associated with the main 
branch. In the next snapshot, the fifth branch then passes within 
R$_{\rm vir,parent}$, and so its mass gets associated with the main branch 
leading to the large increase in mass in snapshots 196 - 200.

Compared to the old \textsc{VELOCIraptor}, the updated \textsc{VELOCIraptor} 
+ \textsc{TreeFrog} along with \textsc{WhereWolf} tracks halos well, but 
there are some large fluctuations in mass in the 3$^{\rm rd}$ and 10$^{\rm th}$ branches even when the halo is a subhalo (so its mass will be exclusive). This 
large fluctuation in mass arises because it is ambiguous as to which 
(sub)halo particles should be assigned to.

Comparison of Figure \ref{fig:VELoldDendoBad} and \ref{fig:VELDendoBad} shows that most of the issues with 
the old \textsc{VELOCIraptor} have been addressed with the updates to \textsc{VELOCIraptor} and \textsc{TreeFrog}, and the implementation of \textsc{WhereWolf}. \textsc{WhereWolf} tracks 
halos until they completely merge with the main-branch; this means that
halos do not suddenly disappear because \textsc{VELOCIraptor} cannot find 
them. However some branches do suffer from rapid changes in mass, even when 
the branch is a subhalo; this is because of the difficulty of tracking the
systems during the multiple massive merger events.

\begin{figure*}[hp]
\centering
 \includegraphics[width=\textwidth]{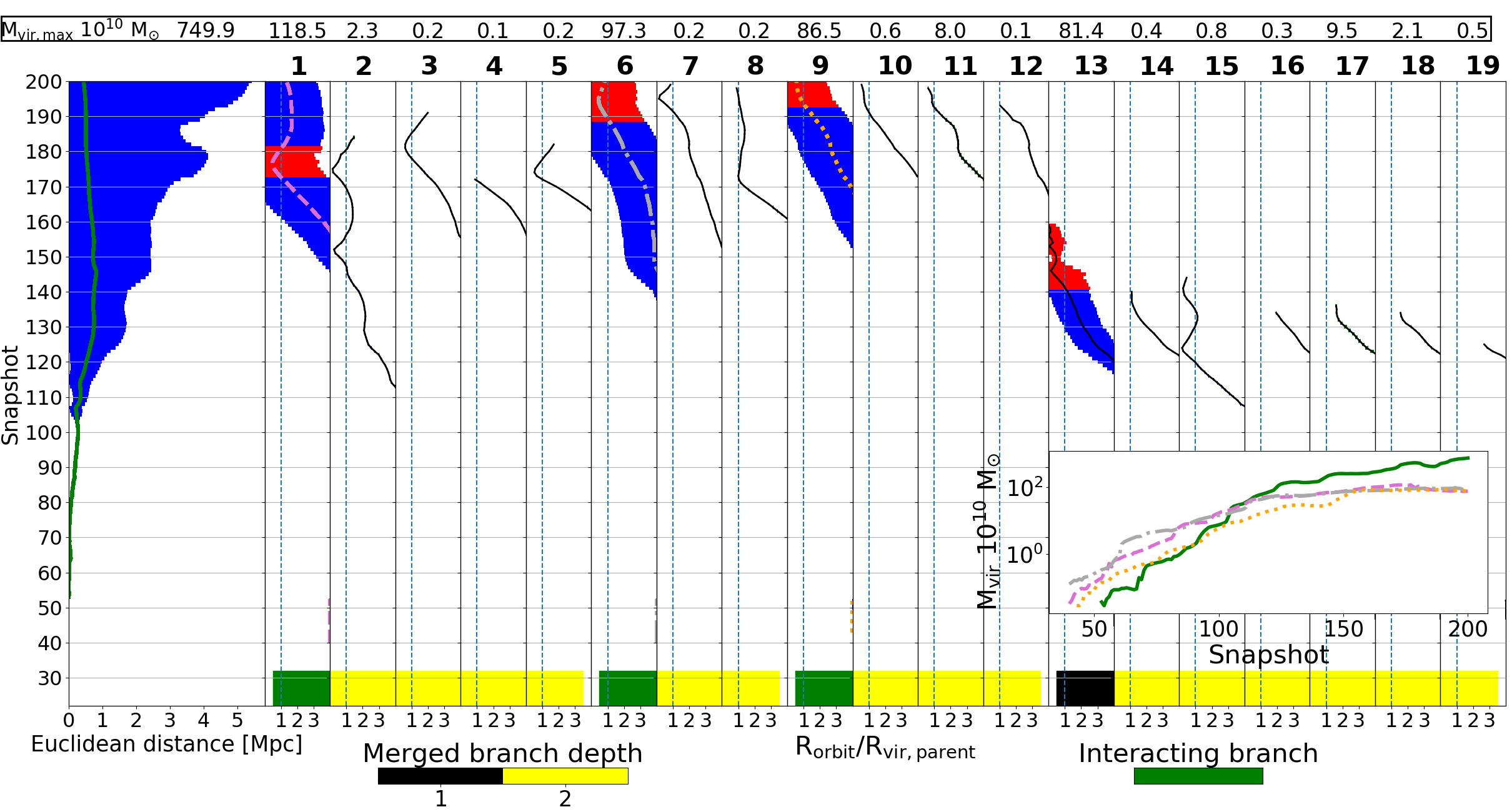}
 \caption{QUAD1 merger tree from \textsc{Rockstar} and \textsc{Consistent Trees}.}  
 \label{fig:RockDendoBad}
\end{figure*}

The dendogram for QUAD1 from the \textsc{Rockstar} and \textsc{Consistent 
Trees} catalogue is shown in Figure \ref{fig:RockDendoBad}. The major 
merging branches have been tracked well and undergo smooth evolution as 
shown in the inset plot.

However, the \textsc{Rockstar} and \textsc{Consistent Trees} algorithm is still not perfect. The 12$^{\rm th}$ branch in Figure \ref{fig:RockDendoBad} does fluctuate in mass as 
it comes into merge, suggesting that \textsc{Rockstar} is unbinding the 
particles too quickly. Particles that are not completely unbound become bound 
again, causing the growth of the halo after snapshot 150. The halo could 
also be picking up some of the loosely bound main-branch halo's particles; 
this is what seems to happen in the third branch of figure \ref{fig:RockgoodMT}. This can lead to problems in a 
SAM where the galaxy inside these halos grow due to their host dark matter 
halo growing; however, because the dark matter mass of the system remains
the same, it leads to an increase in the baryon mass, potentially increasing 
it above the cosmic baryon fraction.

\begin{figure*}[hp]
\centering
 \includegraphics[width=\textwidth]{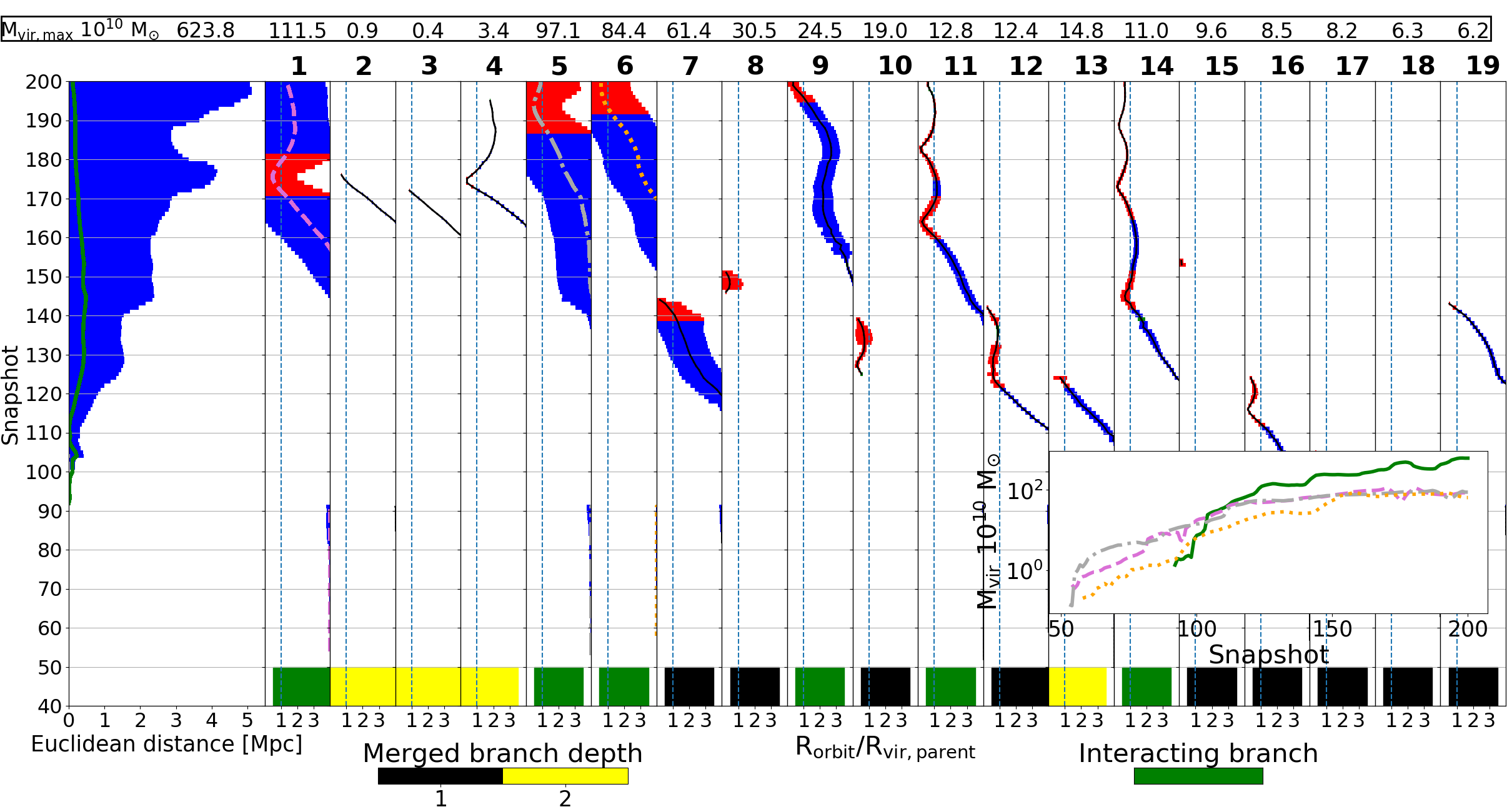}
 \caption{QUAD1 merger tree from the \textsc{AHF} and \textsc{MergerTree} catalogue.}
 \label{fig:AHFDendoBad}
\end{figure*}

The dendogram constructed from \textsc{AHF} + \textsc{MergerTree} is shown 
in Figure \ref{fig:AHFDendoBad}. Overall \textsc{AHF} manages to captures the 
evolution of the large branches coming into merge, where the large change in 
M$_{\rm vir,parent}$ corresponds to one of the large branches passing within 
R$_{\rm vir, parent}$. Nonetheless from the figure it is clear that branches 7 and 8 
suffer from the same problem identified in Figure \ref{fig:AHFgoodMT}, 
where halos that are lost in pericentric passage are found again when it 
starts to exit the halo.

Moreover, branch 10 in Figure \ref{fig:AHFDendoBad} seems to be a continuation of 
branch 13, but it is connected to the wrong halo. This happens because 
the large increase in mass in branch 13 at snapshot 124 
means it has a poor link with the lower mass halo in branch 10 at the next 
snapshot. Furthermore, there is a large exchange of mass between branches 10 and 12, 
where branch 10 hosts the branch 12 temporarily before they merge.

\section{Merger density plots} \label{app:complete}

\begin{figure*}[hp]
\centering
 \includegraphics[width=\textwidth]{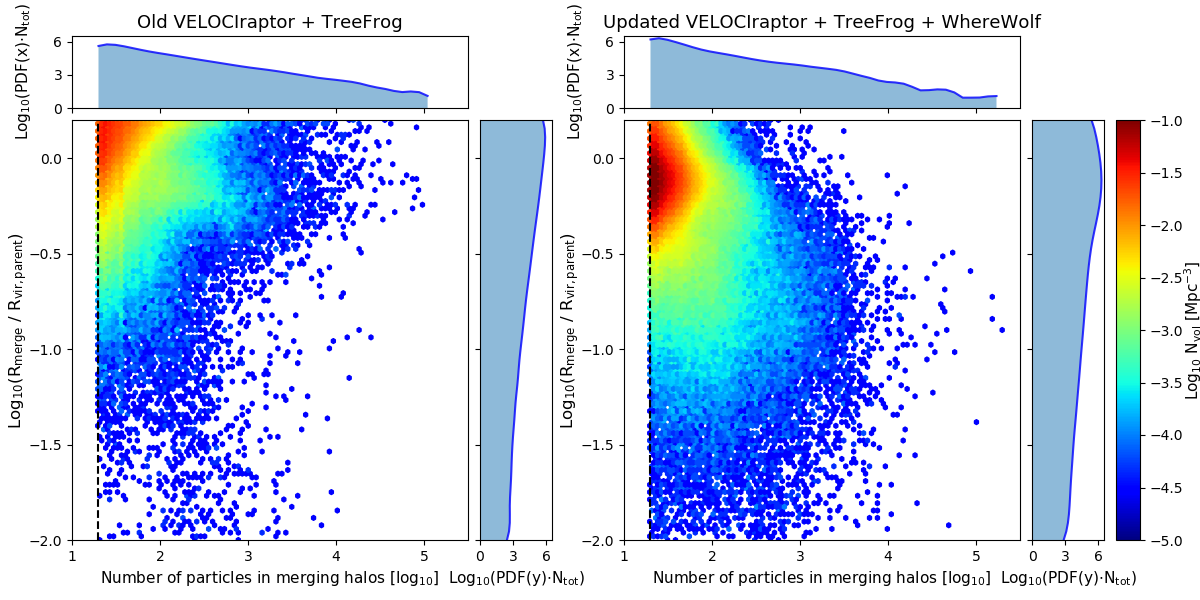}
 \caption{This shows the merger density plots for the old  \textsc{VELOCIraptor} and \textsc{TreeFrog} (left) and the updated \textsc{VELOCIraptor} and \textsc{TreeFrog} catalogues (right). The black dashed line in the plots is for halos with 20 particles, which is the smallest halo stored for all halofinders. The colours represent the log of volumetric counts of the halos. The side plots show the probability density function (PDF), found by using a kernel density estimator \citep{rosenblatt_remarks_1956,parzen_estimation_1962} along each axis, multiplied by the total number of halos present in the figure.   \label{fig:2dhistVEL}}   
\end{figure*}

\begin{figure*}[hp]
\centering
 \includegraphics[width=\textwidth]{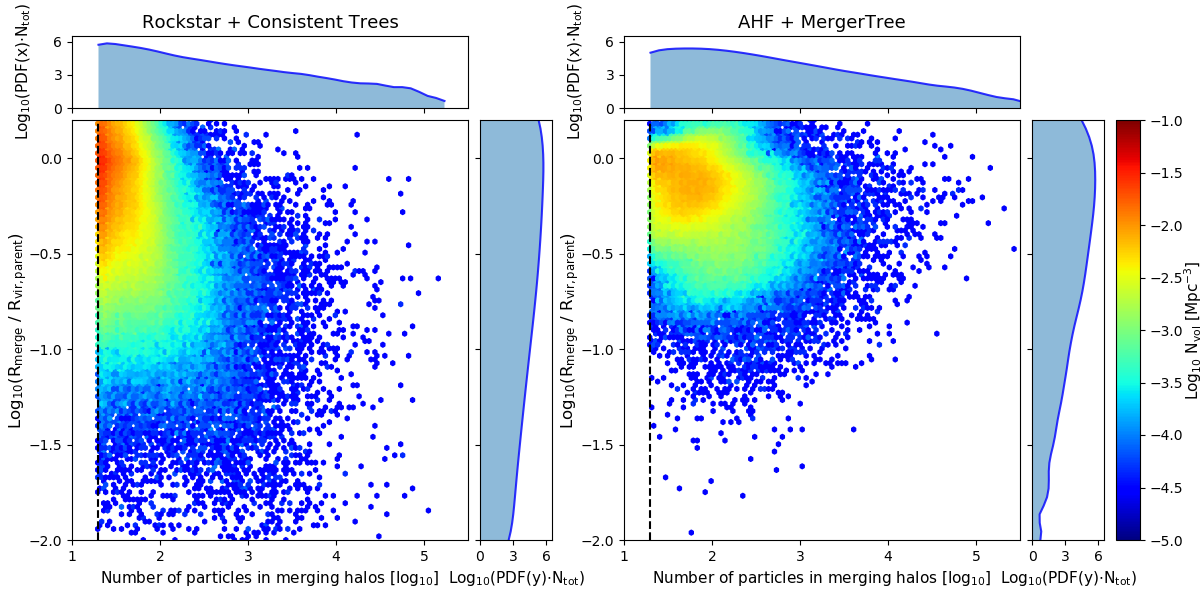}
 \caption{The merger density plots for the \textsc{Rockstar} (left) and \textsc{AHF} (right) catalogues. \label{fig:2dhistRockAHF}}   
\end{figure*}

\begin{figure}[ht]
\centering
 \includegraphics[width=0.46\textwidth]{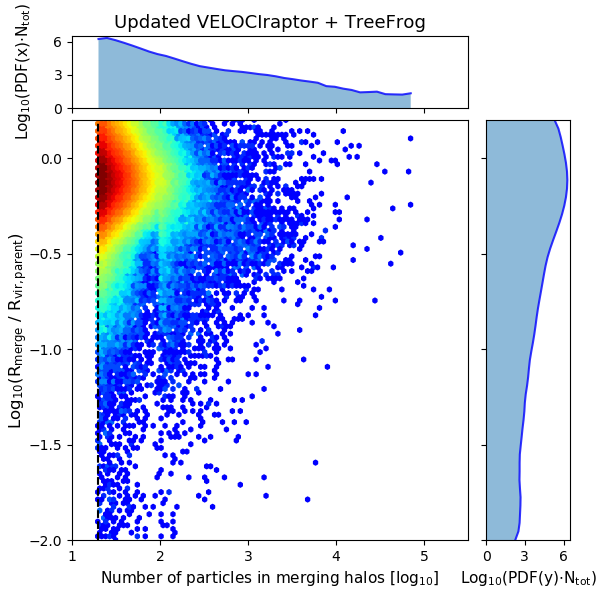}
 \caption{The merger density plots for the updated \textsc{VELOCIraptor} and \textsc{TreeFrog} catalogue, before \textsc{WhereWolf} has been run. \label{fig:2dhistnewVEL}}   
\end{figure}

Another novel way of comparing global merger tree properties of two different catalogues are merger density plots shown in Figures \ref{fig:2dhistVEL}, \ref{fig:2dhistRockAHF} and \ref{fig:2dhistnewVEL}. These plots 
show the distance at which (sub)halos merge with their larger parent halo 
and the number of particles comprising the (sub)halo when it was last found. 
These plots probe completeness of the samples and complement the dendograms. The plots show a 2d histogram of the 
number or particle in merging halos against the ratio of the merger radius 
(R$_{\rm merge}$) to the virial radius of its parent halo it is merging with 
(R$_{\rm vir,parent}$). The colours represent the volumetric counts of halos 
within bins along each axis. 

Ideally, the merging (sub)halos would preferably be in the bottom left hand corner of the merger density plots, where the halos have merged well within the parents R$_{\rm  vir,parent}$ and are left with very few particles. Halos in the top right are halos that have merged with many particles and are far away from their parents R$_{\mathrm{vir,parent}}$, which means that the halo has not been well 
tracked until its disruption.

The merger density plots for the old and updated \textsc{VELOCIraptor} are 
shown in Figure \ref{fig:2dhistVEL}. By studying the old/updated 
\textsc{VELOCIraptor} merger density plots, we can see that the updates have shifted 
the majority of halos to be merged within their parents 
R$_{\rm  vir,parent}$. The addition of  \textsc{WhereWolf} tracking missing halos 
means that halos merge much deeper into their parents R$_{\rm  vir,parent}$. The 
difference can be seen clearly in Figure \ref{fig:2dhistnewVEL}, which shows 
the merger density plot for the updated \textsc{VELOCIraptor} and 
\textsc{TreeFrog} before \textsc{WhereWolf} has been run on it. The 
addition of \textsc{WhereWolf} allows for a more careful and complete
tracking of halo and subhalo orbits.

The merger density plots for both \textsc{AHF} and \textsc{Rockstar} are shown 
in Figure \ref{fig:2dhistRockAHF}. From the plots we can see that \textsc{AHF} struggles to 
identify halos well inside the parents' R$_{\rm vir,parent}$, primarily 
because \textsc{AHF} struggle to identify (sub)halos in overdense 
backgrounds. In contrast \textsc{Rockstar} performs much better because it is a phase 
space halo-finder, like \textsc{VELOCIraptor} and \textsc{Consistent Trees} 
gravitationally evolves the position of the halo once it had been lost. 

By comparing the merger density plots in both Figure \ref{fig:2dhistVEL} and 
\ref{fig:2dhistRockAHF}, it can be seen that, overall, the number density of 
halos is less in the \textsc{AHF} and \textsc{Rockstar}. In the case of 
\textsc{AHF}, it cannot pick out as many objects in dense environments. 
\textsc{Rockstar} does have more than \textsc{AHF} which is due to it being 
a phase space halo finder, but not as many merged (sub)halos as 
\textsc{VELOCIraptor}. This is because \textsc{Consistent Trees} 
gravitationally evolves positions of some (sub)halos for up to four snapshots \citep{behroozi_gravitationally_2013}, which means that if something evaporates within the last four snapshots, it will not merge. In addition the algorithm also removes any subhalo that does not exist for more than 10 snapshots, effectively discarding any subhalo which is fluctuating around the particle limit and also any subhalo that may have undergone fragmentation on infall.

Ideally, we expect halos to slowly lose mass until their pericentres are a small fraction of the virial radius, and only truly vanish when they are near the resolution limit of the simulation. These figures indicate this is not always the case, with some halo-finders and tree-builders implying that halos merge well outside the virial radius of their host while still resolved by thousand of particles. Halos merging while still having a large number of particles and being outside R$_{\rm vir,parent}$ suggests a problem with the codes, but the solution is not immediately apparent from the merger density plots. The dendograms provide a clearer picture of what happens in these situations, therefore presenting possible solutions to these halos merging at large radii.


\section{DISCUSSION}

The merger tree dendogram plots are high information density visualisations of the lives of (sub)halos extracted using halo finders and tree builders. This enables not only a detailed examination of how the halo finders and tree builders are performing, but also for other researchers to find the best merger trees for their desired project. The dendograms provide a comparison tool for the merger tree builders and a novel visualization of what the merger trees look like.

These dendograms have been useful to help identify cases where either the halo finder has not properly identified the halo or tree building algorithm has not correctly connected up the halos across snapshots. In addition the dendogram can be used to identify exactly where the problems arise enabling a much quicker refinement process for either the halo finder or tree builder.

Utilizing these dendograms we hope to address what has been a root-problem within SAMs: how to treat merging satellite galaxies within simulations. While ideally these codes should trace galaxy mergers from first infall to complete coalescence, halo finders and tree builders are not always able to provide such a picture. This comes down to the ability of the halo finder to track halos well inside the virial radius of the host halo \citep[see][for more information]{pujol_nifty_2017}. 

Typically, it is assumed that the satellite galaxy associated with the halo does not merge when its host halo is lost.  SAMs use a few different methods to determine the trajectory and lifetime of these "orphaned" galaxies \citep{guo_dwarf_2011}.  One approach is where the galaxy is merged immediately when its halo has been lost; another approach is where an analytical orbit is determined from when its halo is lost and is continually decreased until it is merged; a further approach is where the trajectory is determined from its most bound particles of its host halo. Some approaches even uses a combination of these methods \citep{pujol_nifty_2017}. These different methods can have varying effects on the abundances of galaxies produced due to the underlying assumptions such as mass loss rate, dynamical friction timescale etc. \cite{robotham_galaxy_2011} demonstrates issues with this approach in modern SAMs finding more satellites on close orbits than exist in the observational GAMA survey data.

While plots like the merger density plots are useful since they indicate the possible presence of "orphaned" galaxies. But it is not clear what is happening in each situation from the plots. Using the dendograms, it will be clear when these sort of events happen and how the halo finder/ tree builder deals with the situation. By highlighting when these events happen  we hope to address this problem and improve merger trees built by these codes.

The dendograms will also be useful when comparing between different halo finders and tree builders. In this paper we have shown the dendograms for three different halo finders and their respective tree building algorithms, giving an idea of how these plots can be used for future comparison projects. This work will also give researchers a better understanding of what to expect when they use merger trees from a particular halo finder. 

We believe that the dendograms will not only be useful for tracking dark matter halos in a simulation, but also for tracking baryonic galaxies. The colour of the points could be changed depending on the requirements, e.g.\ to show stellar/gas mass content. There are many other possibilities, but care should be taken not to trade information density for comprehensibility. When developing the code, we found it was difficult to add much more complexity to the dendograms without compromising their accessibility. That said, the code is hosted on an open source repository in order to encourage community uptake and adaptation. Furthermore, a web interface could be created for the dendograms, whereby clicking on each branch shows another dendogram displaying everything that has merged with it.

\begin{acknowledgements}
We would like to thank Greg B. Poole and Ainulnabilah B. Nasirudin for their clear and constructive comments. RP is supported by a University of Western Australia Scholarship.  AR acknowledges the support of ARC Discovery Project grant DP140100395. CP is supported by ARC Future Fellowship FT130100041. PJE is supported by the Australia Research Council (ARC) Discovery Project Grant DP160102235 and ARC Centre of Excellence ASTRO 3D through project number CE170100013 Part of this research was undertaken on Raijin, the NCI National Facility in Canberra, Australia, which is supported by the Australian commonwealth Government. Parts of this research were conducted by the Australian Research Council Centre of Excellence for All Sky Astrophysics in 3 Dimensions (ASTRO 3D), through project number CE170100013.
\end{acknowledgements}

\begin{appendices}
\renewcommand\thefigure{\thesection.\arabic{figure}}   
\renewcommand\thetable{\thesection.\arabic{table}}    
\setcounter{figure}{0}  

\section{How the dendogram builder code works} \label{app:code}

\begin{figure}
 \includegraphics[width=0.42\textwidth]{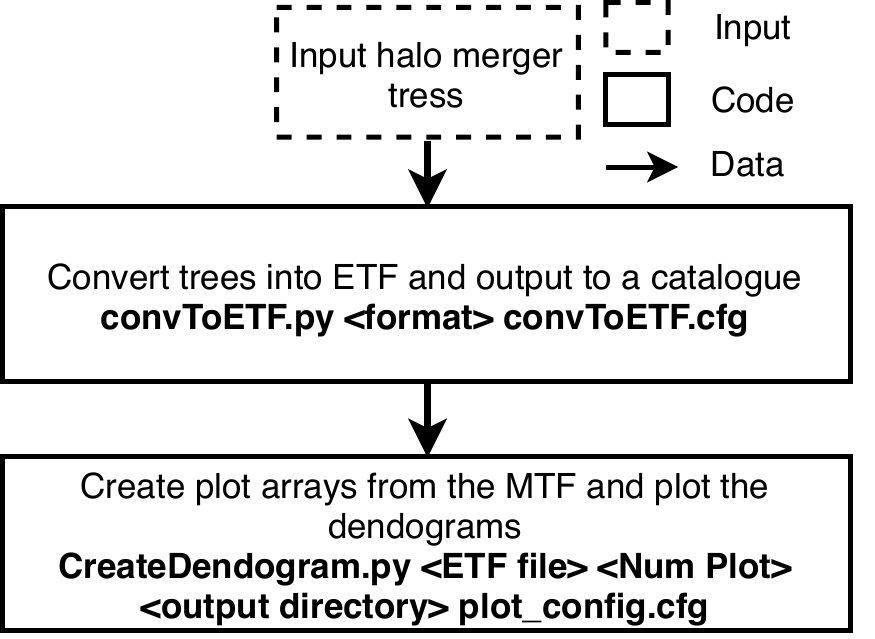}
\caption{ How the dendogram builder code works. Format is the format of the merger tree, convToETF.cfg is the configuration file used to create the ETF from the format given. Num plot is the number of dendograms to plotted, output directory is the directory where the dendograms will be placed and  plot\textunderscore\/config.cfg is a config file which provides all the information for plotting. \label{fig:flowchart}}
\end{figure}

The code requires the tree information be in Efficient Tree Format (ETF), this is where every halo knows the halo that it started the simulation in (Root-Progenitor), the halo which it ended the simulation in (End/Root-Descendant), in addition to knowing where it was in the previous snapshot (Progenitor) and where is going to be in the next snapshot (Descendant). This format makes processing the trees and building the dendograms much faster, once this initial ETF pre-processing is done. The header of the ETF catalogue contains all the required simulation information.

Next, the indexes of the trees which are to be plotted needs to be selected. By default this is done in size order of the halos in the final snapshot. Then  by using the End/Root-Descendant and Root-Progenitors the full merger tree can be extracted for a halo. In addition, using identity of the halo's host a full interaction tree can be built to plot in the dendogram. This data is then used to create the plotting arrays, these are:
\begin{description}
\item[xposData] A N$_{snaps}$ x N$_{branches}$ array of the distance moved for the main branch and radial position for the deeper branches.
\item[sizeData] A N$_{snaps}$ x N$_{branches}$ array of the size of the data points.
\item[colData] A N$_{snaps}$ x N$_{branches}$ array of the colour of the data points.
\item[sortIndx] A 1 x N$_{branches}$ array of the sorted index from the sizeData array for each branch.
\item[branchIndicator] A 1 x N$_{branches}$ array of the index of which branch this branch merges with and also indication if the branch is a subhalo.
\item[depthIndictator] A 1 x N$_{branches}$ array of the temporal depth of each from the main branch.
\end{description}

Where N$_{branches}$ is the number of branches which have a unique Root-Progenitor and the N$_{snaps}$ is the number of snapshots in the simulation. These arrays are then used to plot the dendogram from the specified options in the plotting config file, a sample config file is provided with the publicly available code.

A flow chart summarizing how this code works is shown in Figure \ref{fig:flowchart}, for more details please see the MergerTree-Denograms repository.

\section{Efficient Tree Format (ETF)} \label{app:form}

\begin{table*}[ht]
\caption{ Table showing the minimum amount of data available in ETF for version 1.0. \label{tab:form}}
\centering
\small
\begin{tabular}{p{1cm} p{2cm} p{4cm} p{9cm}}
\hline
Group &Dataset &  Datatype & Comments \\ \hline

/Header & & & \\
 &StartSnap & \texttt{Int(32bit)} attribute & The desired snapshot to start the plotting the dendograms \\
 &EndSnap & \texttt{Int(32bit)} attribute & The desired snapshot to end the plotting the dendograms \\
 &NSnap & \texttt{Int(32bit)} attribute & Number of desired snapshots in the simulation to plot\\
 &h & \texttt{Real(32bit)} attribute & The reduced Hubble parameter h = H$_{0}$/100 in units km/s Mpc$^{-1}$ \\
 &Boxsize & \texttt{Real(32bit)} attribute & The comoving box-size of the simulation in Mpc (no h)\\
 &HALOIDVAL & \texttt{Int(64bit)} attribute & The value which to offset the halo snapshot in the ID to make it temporally unique \\
 &CosmoSim & \texttt{Boolean} attribute & Flag if this is catalogue is from a cosmological simulation False = no, True = yes. \\
 &Munit & \texttt{String} attribute & The mass unit, in ETF this is always 10$^{10}$ M$_{\odot}$ \\
 &Lunit & \texttt{String} attribute & The length unit, in ETF this is always Mpc \\
&Vunit & \texttt{String} attribute & The velocity unit, in ETF this is always km/s \\
&... & ... & Additional header information if desired, if a cosmological simulation the cosmological parameters are suggested. \\
\hline
/Snap\_\# & & & Each dataset within this group has an attribute stating its original name from the catalogue it was converted from if it existed. \\
&Redshift|Time & \texttt{Real(32bit)} attribute & The redshift if a cosmological simulation or time if not, for this snapshot \\
&StartProgenitor & \texttt{Int(32bit)}|\texttt{Int(64bit)} & ID of the halo when it first formed \\
&Progenitor & \texttt{Int(32bit)}|\texttt{Int(64bit)} & ID of the halos progenitor  \\
&Descendant & \texttt{Int(32bit)}|\texttt{Int(64bit)} & ID of the halos descendant  \\
&EndDescendant & \texttt{Int(32bit)}|\texttt{Int(64bit)} & ID of the halo when it evaporated or at snapshot /Header/endSnap \\
&HaloID & \texttt{Int(32bit)}|\texttt{Int(64bit)} & The temporally unique ID for this halo \\
&HostHaloID & \texttt{Int(32bit)}|\texttt{Int(64bit)} & ID of the the host of the (sub)halo, -1 if it has no host  \\
&Pos & \texttt{Real(32bit)} array(N$_{\rm halo}$,3) & The comoving position of the halo in the simulation in Mpc \\
&Vel &  \texttt{Real(32bit)} array(N$_{\rm halo}$,3) & The physical velocity of the halo in the simulation in km/s \\
&Mass & \texttt{Real(32bit)} & User definable mass off the halo in units of 10$^{10}$M$_{\odot}$ \\
&Radius & \texttt{Real(32bit)} & User definable radius off the halo in units of  Mpc \\
&...& ... & Addtional data to use for setting the size or the colour of the points on the plot \\

\end{tabular}
\end{table*}

Table \ref{tab:form} shows the required format of the input HDF5 file to create the dendogram plots, and Figure \ref{fig:ETFsch} shows a diagram of the merger tree information available in ETF format. With the header containing the required information for the simulation. The IDs in the format are given by:
\noindent
\begin{equation*}
\mathrm{HaloID} = \mathrm{snapshot} \  \times \ \mathrm{HALOIDVAL} \ + \ i_{\mathrm{halo}} \ + \ 1, 
\end{equation*}
where $i_{\rm halo}$ is the index of the halo within the current snapshot. The IDs follow this format as this enables a quick parsing of the tree to create the plotting arrays needed to build described in appendix \ref{app:code} to create the dendograms. 

\begin{table}
\centering
\caption{ Timing to convert different formats into ETF. \label{tab:timings}}
\begin{tabular}{cc}
Format & Time to convert (CPU hrs) \\ \hline
\textsc{VELOCIraptor} & <0.1 \\
\\
\textsc{Rockstar}  & 1.5 \\
\\
\textsc{AHF}  & 20 \\ \hline

\end{tabular}

\end{table}

This format is required since it enables a quick traversal of the halo merger trees, enabling for a much quicker building of the merger tree for the dendogram. Table \ref{tab:timings} shows the timings to convert the \textsc{VELOCIraptor}, \textsc{Rockstar} and \textsc{AHF} format into ETF. For the halo catalogue built using the 40Mpc/h 512$^{3}$ particle SURFS simulation. This format is very similar to the \textsc{VELOCIraptor} (Elahi et al. in prep), with only field name different hence why the conversion takes virtually no time. Where as \textsc{Rockstar} and \textsc{AHF} IDs do not contain information such as the halo snapshot and the index (where they are located in the snapshot catalogue). Both \textsc{VELOCIraptor} and \textsc{Rockstar} have catalogues that contain the merger tree along with the halo properties whilst \textsc{AHF} does not. This means for \textsc{AHF} it has to be found in both the halo and merger tree catalogues.   \textsc{AHF} (\textsc{MergerTree}) does not have an efficient way of reporting whether a halo has no progenitors, except if it does not exist in the catalogue. These reasons are why \textsc{AHF} take longer than any of the other formats to convert to ETF, but a parallel conversion tool has been built to convert AHF to reduce the real time taken to convert.

\section{Example exclusive mass dendogram }
 \label{app:exclMass}

\begin{figure*}[ht]
\centering
 \includegraphics[width=\textwidth]{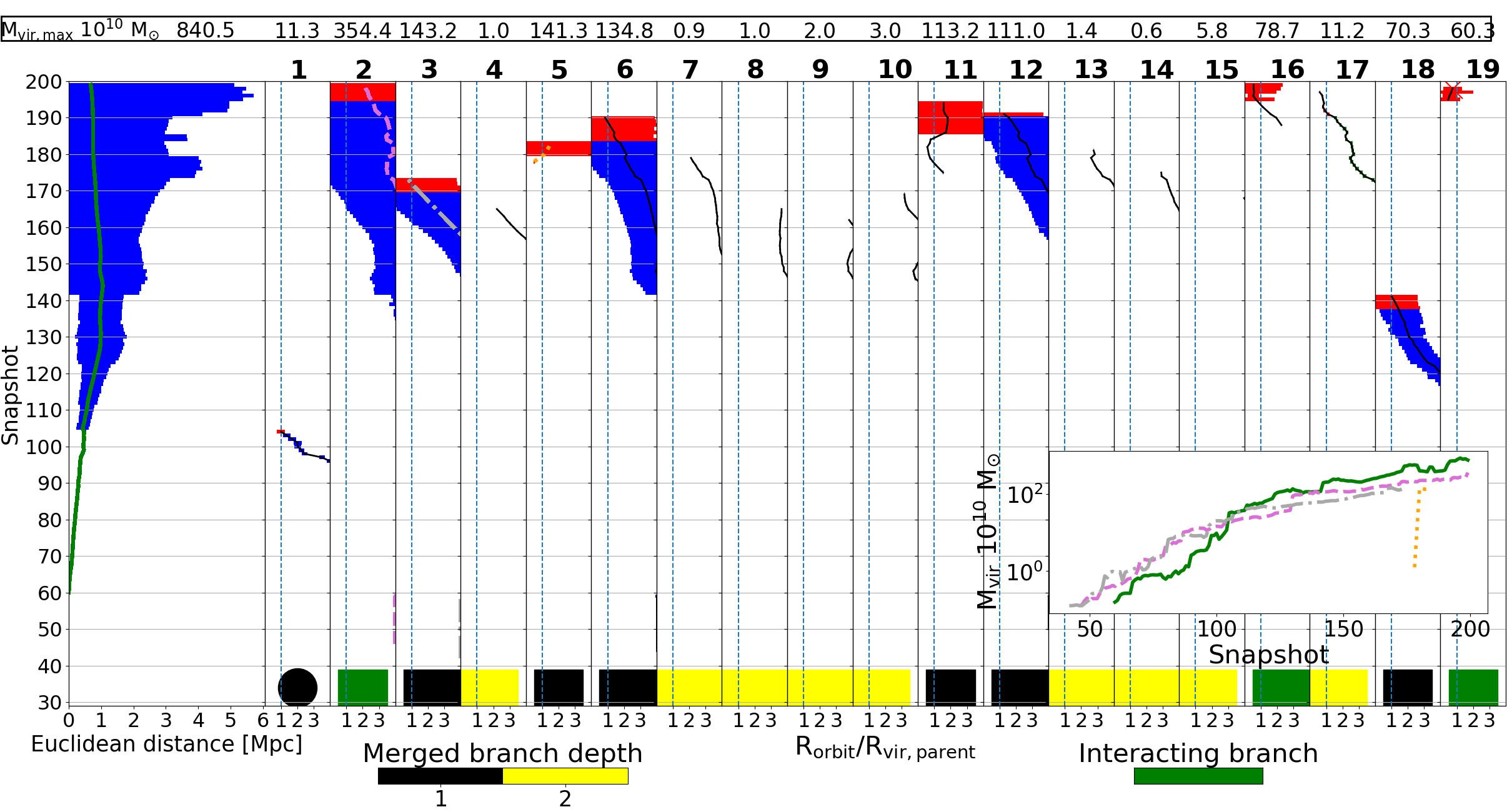}
 \caption{QUAD1 merger tree from updated \textsc{VELOCIraptor} + \textsc{TreeFrog} catalogue where the masses are exclusive, before \textsc{WhereWolf} has been run.}   \label{fig:VELmassExcl}
\end{figure*}

This example shows the same case as in Figure \ref{fig:VELDendoBad} but where the mass is exclusive, so the mass for a halo is just the mass of the halo itself and does not contain the masses  of any subhalos of the halo. The changes in parent halo mass is much smoother than for the inclusive dendogram. This is due to it not being affected as much when halos become a subhalo of the parent halo. The dendogram is also shown before \textsc{WhereWolf} was run on the updated \textsc{VELOCIrator} and \textsc{TreeFrog} catalogue so the improvement can clearly be seen. From Figure \ref{fig:VELDendoBad}, it can be seen that \textsc{WhereWolf} is able to connect up the halo that was assigned to the wrong branch in the fifth branch due to the halo missing for over 10 snapshots. 

From Figure \ref{fig:VELmassExcl}, it can be seen that \textsc{VELOCIraptor} merges most of the large branches with the main branch since it can no longer track them. In comparison, \textsc{WhereWolf} is able to continue track these branches and accurately reconstruct their masses until they are completely dispersed or the simulation ends. This can vastly change the evolution of the main branch and what happens to its central galaxy.

\end{appendices}
\bibliographystyle{pasa-mnras}
\bibliography{Zotero}

\end{document}